\documentclass[12pt]{article}
\usepackage{fancyhdr}

\usepackage{mathrsfs}
\usepackage[T1]{fontenc}
\usepackage{setspace}
\usepackage{amsfonts}
\usepackage{amssymb}
\usepackage{amsmath}
\usepackage{epsfig}
\usepackage{latexsym}
\usepackage{color}
\usepackage{graphicx}
\usepackage{nicefrac}
\usepackage[latin1]{inputenc}
\usepackage{slashed}
\usepackage{multirow}
\usepackage{comment}
\usepackage{soul}
\usepackage{hyperref}
\usepackage{cite}
\usepackage{slashed}
\usepackage{simpler-wick}
\usepackage{array}
\usepackage{tabu}

\usepackage[titletoc]{appendix}

\def\hybrid{\topmargin -20pt    \oddsidemargin 0pt
        \headheight 0pt \headsep 0pt
        \textwidth 6.25in       
        \textheight 9.25in       
        \marginparwidth .875in
        \parskip 5pt plus 1pt   \jot = 1.5ex}

\hybrid

\def\baselinestretch{1.2}

\catcode`\@=11

\def\marginnote#1{}
\newcount\hour
\newcount\minute
\newtoks\amorpm
\hour=\time\divide\hour by60
\minute=\time{\multiply\hour by60 \global\advance\minute by-\hour}
\edef\standardtime{{\ifnum\hour<12 \global\amorpm={am}%
        \else\global\amorpm={pm}\advance\hour by-12 \fi
        \ifnum\hour=0 \hour=12 \fi
        \number\hour:\ifnum\minute<10 0\fi\number\minute\the\amorpm}}
\edef\militarytime{\number\hour:\ifnum\minute<10 0\fi\number\minute}

\def\draftlabel#1{{\@bsphack\if@filesw {\let\thepage\relax
   \xdef\@gtempa{\write\@auxout{\string
      \newlabel{#1}{{\@currentlabel}{\thepage}}}}}\@gtempa
   \if@nobreak \ifvmode\nobreak\fi\fi\fi\@esphack}
        \gdef\@eqnlabel{#1}}
\def\@eqnlabel{}
\def\@vacuum{}
\def\draftmarginnote#1{\marginpar{\raggedright\scriptsize\tt#1}}

\def\draft{\oddsidemargin -.5truein
        \def\@oddfoot{\sl preliminary draft \hfil
        \rm\thepage\hfil\sl\today\quad\militarytime}
        \let\@evenfoot\@oddfoot \overfullrule 3pt
        \let\label=\draftlabel
        \let\marginnote=\draftmarginnote
   \def\@eqnnum{(\theequation)\rlap{\kern\marginparsep\tt\@eqnlabel}%
\global\let\@eqnlabel\@vacuum}  }

\def\preprint{\twocolumn\sloppy\flushbottom\parindent 2em
        \leftmargini 2em\leftmarginv .5em\leftmarginvi .5em
        \oddsidemargin -.5in    \evensidemargin -.5in
        \columnsep .4in \footheight 0pt
        \textwidth 10.in        \topmargin  -.4in
        \headheight 12pt \topskip .4in
        \textheight 6.9in \footskip 0pt
        \def\@oddhead{\thepage\hfil\addtocounter{page}{1}\thepage}
        \let\@evenhead\@oddhead \def\@oddfoot{} \def\@evenfoot{} }

\def\numberbysection{\@addtoreset{equation}{section}
        \def\theequation{\thesection.\arabic{equation}}}

\def\underline#1{\relax\ifmmode\@@underline#1\else
        $\@@underline{\hbox{#1}}$\relax\fi}

\def\titlepage{\@restonecolfalse\if@twocolumn\@restonecoltrue\onecolumn
     \else \newpage \fi \thispagestyle{empty}\c@page\z@
        \def\thefootnote{\fnsymbol{footnote}} }

\def\endtitlepage{\if@restonecol\twocolumn \else \newpage \fi
        \def\thefootnote{\arabic{footnote}}
        \setcounter{footnote}{0}}  

\catcode`@=12
\relax

\def\figcap{\section*{Figure Captions\markboth
        {FIGURECAPTIONS}{FIGURECAPTIONS}}\list
        {Figure \arabic{enumi}:\hfill}{\settowidth\labelwidth{Figure
999:}
        \leftmargin\labelwidth
        \advance\leftmargin\labelsep\usecounter{enumi}}}
 \relax
\def\tablecap{\section*{Table Captions\markboth
        {TABLECAPTIONS}{TABLECAPTIONS}}\list
        {Table \arabic{enumi}:\hfill}{\settowidth\labelwidth{Table
999:}
        \leftmargin\labelwidth
        \advance\leftmargin\labelsep\usecounter{enumi}}}
 \relax
\def\reflist{\section*{References\markboth
        {REFLIST}{REFLIST}}\list
        {[\arabic{enumi}]\hfill}{\settowidth\labelwidth{[999]}
        \leftmargin\labelwidth
        \advance\leftmargin\labelsep\usecounter{enumi}}}
 \relax
\makeatletter
\newcounter{pubctr}
\def\publist{\@ifnextchar[{\@publist}{\@@publist}}
\def\@publist[#1]{\list
        {[\arabic{pubctr}]\hfill}{\settowidth\labelwidth{[999]}
        \leftmargin\labelwidth
        \advance\leftmargin\labelsep
        \@nmbrlisttrue\def\@listctr{pubctr}
        \setcounter{pubctr}{#1}\addtocounter{pubctr}{-1}}}
\def\@@publist{\list
        {[\arabic{pubctr}]\hfill}{\settowidth\labelwidth{[999]}
        \leftmargin\labelwidth
        \advance\leftmargin\labelsep
        \@nmbrlisttrue\def\@listctr{pubctr}}}
 \relax
\makeatother
\newskip\humongous \humongous=0pt plus 1000pt minus 1000pt

\newif\ifdtup

\relax

\def\be{\begin{equation}}
\def\ee{\end{equation}}
\def\ba{\begin{eqnarray}}
\def\ea{\end{eqnarray}}

\def\del{\partial}

\def\a{\alpha}

\def\b{\beta}

\def\d{\delta}

\def\p{\pi}

\def\m{\mu}
\def\n{\nu}

\def\l{\lambda}

\def\s{\sigma}

\def\no{\noindent}

\def\IR{\relax{\rm I\kern-.18em R}}

\def\IR{\relax{\rm I\kern-.18em R}}
\def\IL{\relax{\rm I\kern-.18em L}}

\def\inv{^{\raise.15ex\hbox{${\scriptscriptstyle -}$}\kern-.05em 1}}

\begin{document}

\renewcommand{\theequation}{\thesection.\arabic{equation}}
\csname @addtoreset\endcsname{equation}{section}

\newcommand{\beq}{\begin{equation}}
\newcommand{\eeq}[1]{\label{#1}\end{equation}}
\newcommand{\ber}{\begin{equation}}
\newcommand{\eer}[1]{\label{#1}\end{equation}}
\newcommand{\eqn}[1]{(\ref{#1})}
\begin{titlepage}
\begin{center}

${}$
\vskip .2 in

{\Large\bf  Classical solutions  of  $\l$-deformed coset models}

\vskip 0.4in

{\bf Dimitrios Katsinis${}^{1,2}$ and Pantelis Panopoulos${}^{1}$}
\vskip 0.1in

\vskip 0.1in

 {\em${}^{1}$
Department of Nuclear and Particle Physics,\\
Faculty of Physics, National and Kapodistrian University of Athens,\\
Athens 15784, Greece\\
}
{\em${}^{2}$ Instituto de F\'isica, Universidade de S\~ao Paulo,\\
 Rua do Mat\~ao Travessa 1371, 05508-090 S\~ao Paulo, SP, Brazil}

\vskip 0.1in

{\footnotesize  \texttt { email: dkatsinis@phys.uoa.gr,  ppanopoulos@phys.uoa.gr}}

\today

\vskip .5in
\end{center}

\centerline{\bf Abstract}

\no

We obtain classical solutions of $\l$-deformed $\s$-models based on $SL(2,\mathbb{R})/U(1)$ and $SU(2)/U(1)$ coset manifolds. Using two different sets of coordinates, we derive two distinct classes of solutions. The first class  is expressed in terms of hyperbolic and trigonometric functions, whereas the second one in terms of elliptic functions. We analyze their properties along with the boundary conditions and discuss string systems that they describe. It turns out that there is an apparent similarity between the solutions of the second class and the motion of a pendulum.

\vskip .4in
\noindent
\end{titlepage}
\vfill
\eject

\newpage

\tableofcontents

\noindent

\def\baselinestretch{1.2}
\baselineskip 20 pt
\noindent



\setcounter{equation}{0}

\renewcommand{\theequation}{\thesection.\arabic{equation}}

\newpage

\section{Introduction}

Classical solutions is one of the main subjects in Quantum Field Theory and play an important role in studying its dynamics. A large class of such configurations are the so-called solitons \cite{Tong:2005un}. In general, solitons are solutions which retain their shape, as they propagate at constant velocity. Usually such lumps are topological in nature and saturate a BPS bound. The most studied example, the kinks, interpolate between different classical vacua of the theory. Such solution require the presence of a potential, so that the theory has a non-trivial vacuum structure. These configurations are excitations of the theory, which even though they are not elementary, they exhibit a particle-like behaviour. Kinks can be made to scatter and even create bound states, which are called breathers. An appealing fact about these objects is that the equations describing them are obtained using topological arguments or exploiting integrability and not by attacking the equations of motion straight-forwardly. The reason is the high non-linearity of the equations of motion, which makes the derivation of their solution practically impossible. Having classical solutions one can utilise them in studying several aspects of QFT. These include semi-classical quantization, the vacuum structure of the theory and the effective theory around a non-trivial background solution.
 
In this work, focusing on 2-dimensional field theories, we obtain classical solutions of an integrable class of deformations of WZW-models, namely the $\l$-deformations introduced in \cite{Sfetsos:2013wia}. The generalization  of these models for symmetric spaces is  given in  \cite{Hollowood:2014rla}, while the asymmetric gauging is constructed in \cite{Driezen:2019ykp}, building on  the setup of \cite{Georgiou:2016urf}.   Recently, a lot of attention has been paid in various aspects of these theories. The reason is that they also possess non-perturbative duality symmetries, enabling the exact calculations of various set of observables among which $\b$-functions, $C$-functions and anomalous dimensions for a large class of single and composite operators (see \cite{Itsios:2014lca,Georgiou:2015nka,Georgiou:2018vbb,Georgiou:2019jcf} and references therein).  

It would be interesting though, to move one step further and obtain explicit expressions for classical solutions in order to enlarge the knowledge about the dynamics of such theories. Solving directly the equations of motion is a very hard task due to the high non-linearity. To overcome this, we study $\l$-deformations on coset manifolds $SL(2,\mathbb{R})/U(1)$ and $SU(2)/U(1)$. Even though the non-linearity is still present and the equations of motion are not drastically simplified, there is a way to proceed. In the case of non-linear sigma models having 2-dimensional target space, essentially solving the equations corresponding to the conservation of the Energy-Momentum tensor is equivalent to solving the equations of motion. There is a small caveat it this equivalence but it does not affect our results. In any case, exploiting this remark, along with the fact that these equations are first order and much simpler than the equations of motion, we obtain two different classes of solutions the models under consideration.  It is not clear how they fit in the spectrum of the theory, since our results are not based on topological arguments. 

The structure of the paper is as follows: in Section \ref{sec:general_setup} we present the specific models under study emphasizing the relations between them via analytic continuations. These relations are useful when deriving solutions of the same class. In Section \ref{sec:classical_solutions} we present the method for deriving the solutions of this work. By solving the Energy-Momentum tensor conservation equations we obtain two classes of solutions for vectorially and axially gauged $SL(2,\mathbb{R})/U(1)$ coset model, as well as for the $SU(2)/U(1)$ model. In Section \ref{sec:properties}  We present the admissible boundary conditions and possible brane configurations associated to them. We also analyse the effects of the non-perturbative dualities on the solutions. Section \ref{sec:Discussion} contains conclusions and future directions. Finally, let us mention that Appendix \ref{sec:Jacobi_review} contains a brief review about Jacobi's elliptic functions. The properties of these functions are required in order to understand properties of the second class of solutions. Our main results are gathered in Tables \ref{tab:solutions_sl2_vector} - \ref{tab:solutions_2nd_kink}.

\section{$\lambda$-deformations on coset manifolds}
\label{sec:general_setup}
In this section we introduce the models we are focused on, namely the $SL(2,\mathbb{R})/U(1)$ and $SU(2)/U(1)$ $\l$-deformed models. The vectorially gauged WZW $SU(2)/U(1)$ $\lambda$-deformed model has been worked out in \cite{Sfetsos:2013wia}. In \cite{Sfetsos:2014cea} its relation to the non-compact variant, i.e. the $SL(2,\mathbb{R})/U(1)$ $\lambda$-deformed model, was presented. The axial gauging of these models was constructed in \cite{Driezen:2019ykp}. Some additional details on the parametrization of the models are given in Appendix \ref{sec:parametrization}.

\subsection{$SU(2)/U(1)$ Vector Gauging}

As a first example we present the $SU(2)/U(1)$ $\lambda-$deformed WZW model in the case of vector gauging\footnote{As the group $SU(2)$ does not possess any outer automorphisms, the axial gauging is equivalent to a field re-definition of the vector gauged model, namely $\theta\rightarrow\pi/2-\theta$.}. The action  reads
\begin{equation}
\label{lWitten_su2}
S=\frac{k}{\p}\int\,d^2\s\,e^{2\Phi}\left(\frac{1+\l}{1-\l}\,\del_+y_1\,\del_-y_1+\frac{1-\l}{1+\l}\,\del_+y_2\,\del_-y_2\right),\qquad e^{-2\Phi}=1-y_1^2-y_2^2\,\,\,.
\end{equation}
Coordinates $y_1$ and $y_2$ are related to the usual $(\theta,\phi)$ coordinates by
\be
\label{apcoor_su2}
y_1=\cos \theta \cos\phi,\phantom{000}y_2=\cos\theta \sin\phi.
\end{equation}
It is evident that the coordinates $y_1$ and $y_2$ satisfy $y_1^2+y_2^2\leq 1$. For completeness, we also write down equations of motion, which read
\begin{align}
&\del_+\del_-y_1=\frac{y_1\left(\del_+y_1\del_-y_1-\left(\frac{1-\l}{1+\l}\right)^2\del_+y_2\del_-y_2\right)+y_2\left(\del_-y_1\del_+y_2+\del_+y_1\del_-y_2\right)}{y_1^2+y_2^2-1},\label{eom1_su2}\\
&\del_+\del_-y_2=\frac{y_2\left(\del_+y_2\del_-y_2-\left(\frac{1+\l}{1-\l}\right)^2\del_+y_1\del_-y_1\right)+y_1\left(\del_-y_2\del_+y_1+\del_+y_2\del_-y_1\right)}{y_1^2+y_2^2-1}.\label{eom2_su2}
\end{align}
Note that the action \eqref{lWitten_su2}, including the $SL(2,\mathbb{R})$ models presented below,  respects the duality symmetries
\be
\label{duality}
\l\to 1/\l\,\,\,\,,\phantom{0000}k\to-k
\ee
and also
\be
\label{-lsymmetry}
\,\,\l\to-\l\,\,\,\,,\phantom{0000}y_1\leftrightarrow y_2\,\,.
\ee
These symmetries, if enforced on the solutions, imply specific transformations of the configurations' parameters. We will come back to this later on. 

The action \eqn{lWitten_su2} is also written in a new conformally flat form as
\be
\label{eq:action_su2_chi_psi}
S=\frac{k}{\p}\int\,d^2\s\,e^{2\tilde\Phi}\left(\del_+\chi\del_-\chi+\del_+\psi\del_-\psi\right),\qquad e^{-2\tilde\Phi}=e^{-2\chi}-\left(\frac{1+\l}{1-\l}-\frac{4\l}{1-\l^2}\cos^2\psi\right),
\end{equation}
where the coordinates $\chi$ and $\psi$ are defined in term of $\theta$ and $\phi$ by equations \eqref{eq:def_chi_su2} and \eqref{eq:def_psi}. These coordinates are valued in an appropriate domain, so that the dilaton is real and the conformal factor $e^{-2\tilde\Phi}$ positive.

\subsection{$SL(2,\mathbb{R})/U(1)$ Vector Gauging}

Starting with the $SU(2)/U(1)$ model and applying the following transformations
\begin{equation}\label{eq:analytic_continuation}
k\rightarrow -k,\qquad \kappa\rightarrow i\kappa,\qquad \theta\rightarrow i\rho.
\end{equation}
we obtain the $SL(2,\mathbb{R})/U(1)_V$ $\lambda$-deformed model \footnote{From now on we use this abbreviation for the vectorially gauged $SL(2,\mathbb{R})/U(1)$ model and $SL(2,\mathbb{R})/U(1)_A$ for the axially gauged model presented below.}. Of course, the level $k$ ceases being quantized, but this does not concern our analysis. The  transformation of $\kappa$ is required in order to leave the coupling $\lambda$ invariant. Also, transformation of $\theta$ amounts to exchanging the cosines to hyperbolic cosines. Notice that, the overall sign of the action (due to the transformation of $k$) is absorbed by the dilaton, so that the later remains manifestly positive.
In this case, the action reads
\begin{equation}
\label{lWitten_sl2_vector}
S=\frac{k}{\p}\int\,d^2\s\,e^{2\Phi}\left(\frac{1+\l}{1-\l}\,\del_+y_1\,\del_-y_1+\frac{1-\l}{1+\l}\,\del_+y_2\,\del_-y_2\right),\qquad e^{-2\Phi}=y_1^2+y_2^2-1,
\end{equation}
where the coordinates $y_1$ and $y_2$ are related to the coordinates $\rho$ and $\phi$ as 
\be
\label{apcoor_sl2_vector}
y_1=\cosh \rho \cos\phi,\phantom{000}y_2=\cosh \rho\sin\phi.
\end{equation}
The equations of motion coincide with \eqref{eom1_su2} and \eqref{eom2_su2}, since the actions \eqref{lWitten_su2} and \eqref{lWitten_sl2_vector} as functionals of $y_1$ and $y_2$, differ only by an overall sign. Nevertheless, in this case the coordinates $y_1$ and $y_2$ are defined in the complimentary region satisfying $y_1^2+y_2^2\geq 1$. Following the steps of the previous section, the action \eqref{lWitten_sl2_vector} can be rewritten as
\be
\label{eq:action_sl2_chi_psi_vector}
S=\frac{k}{\p}\int\,d^2\s\,e^{2\tilde\Phi}\left(\del_+\chi\del_-\chi+\del_+\psi\del_-\psi\right),\qquad e^{-2\tilde\Phi}=\frac{1+\l}{1-\l}-\frac{4\l}{1-\l^2}\cos^2\psi-e^{-2\chi},
\end{equation}
where the coordinates $\chi$ and $\psi$ are defined in terms of $\rho$ and $\phi$ by  \eqref{chipsi1_ap} and \eqref{eq:def_psi}.

\subsection{$SL(2,\mathbb{R})/U(1)$ Axial Gauging}

Implementing the transformations $\rho\rightarrow \rho+i \pi/2$, which is equivalent to
\begin{equation}\label{vector_to_axial}
y_I\rightarrow i y_I, \qquad\thickspace I=1,2,
\end{equation}
on $SL(2,\mathbb{R})/U(1)_V$ model we obtain the$SL(2,\mathbb{R})/U(1)_A$ $\lambda$-deformed model. Interestingly enough, the transformation, relating  the axial and vector gauged WZW models at full quantum level \cite{Dijkgraaf:1991ba}, also relates the $\lambda-$deformed theories (at least semi-classically). The corresponding action reads
\be
\label{lWitten_sl2_axial}
S=\frac{k}{\p}\int\,d^2\s\,e^{2\Phi}\left(\frac{1+\l}{1-\l}\,\del_+y_1\,\del_-y_1+\frac{1-\l}{1+\l}\,\del_+y_2\,\del_-y_2\right),\qquad e^{-2\Phi}=y_1^2+y_2^2+1,
\end{equation}
where the coordinates $y_1$ and $y_2$ are related to the coordinates $\rho$ and $\phi$ as 
\be\label{apcoor_sl2_axial}
y_1=\sinh \rho \cos\phi,\qquad y_2=\sinh \rho\sin\phi\,\,.
\end{equation}
The equations of motion are obtained by applying the transformation \eqref{vector_to_axial} on the equations of motion \eqref{eom1_su2} and \eqref{eom2_su2}. It amounts to setting $-1$ to $+1$ on the denominators of the right-hand-sides. The action \eqref{lWitten_sl2_axial} can be rewritten as
\be
\label{eq:action_sl2_chi_psi_axial}
S=\frac{k}{\p}\int\,d^2\s\,e^{2\tilde\Phi}\left(\del_+\chi\del_-\chi+\del_+\psi\del_-\psi\right),\qquad e^{-2\tilde\Phi}=\frac{1+\l}{1-\l}-\frac{4\l}{1-\l^2}\cos^2\psi+e^{-2\chi},
\end{equation}
where the coordinates $\chi$ and $\psi$ are defined in terms of $\rho$ and $\phi$ by equations \eqref{eq:def_chi_sl2_ax} and \eqref{eq:def_psi}. Finally, notice that the action \eqref{eq:action_sl2_chi_psi_axial} is obtain by \eqref{eq:action_sl2_chi_psi_vector} by sending  $\chi\rightarrow \chi+i\pi/2$.

\section{Classical Solutions}
\label{sec:classical_solutions}

In this section we derive two different classes of solutions for each model. The first class of solutions is derived using the actions in terms of the coordinates $y_1$ and $y_2$. For the second class of solutions we use the actions in terms of the coordinates $\chi$ and $\psi$. Even though we could have used a different ansatz to solve the equations of motion of the same action, i.e. the one in terms of $y_1$ and $y_2$, we employ the action in terms of $(\chi,\psi)$, which motivates the ansatz more naturally. 

The first class of solutions, corresponding to the actions \eqref{lWitten_su2}, \eqref{lWitten_sl2_vector} and \eqref{lWitten_sl2_axial}, is expressed in terms of hyperbolic-trigonometric functions, whereas the second one, corresponding to the actions \eqref{eq:action_su2_chi_psi}, \eqref{eq:action_sl2_chi_psi_vector} and \eqref{eq:action_sl2_chi_psi_axial}, in terms of Jacobi elliptic functions. As expected, the relations of the models via analytic continuation, are also reflected to the classical solutions of the same type  as well. Recall that \eqref{vector_to_axial} relates the axially and vectorially gauged $SL(2,\mathbb{R})/U(1)$ models, while the $SU(2)/U(1)$ and the vectorially gauged $SL(2,\mathbb{R})/U(1)$ model merely differ by the fact that the coordinates are valued in complementary regions. Therefore, solving one model is enough for solving all of them and  all differences  of the solutions are highlighted wherever it is necessary.

\subsection{A remark on  the solution of  the equations of motion}

It is evident that solving directly the equations of motion \eqref{eom1_su2} is quit difficult. To untie our hands, a trick is needed which will allows us to sidestep this obstacle. Let us consider a non-linear sigma model (we do not include $B_{\m\n}$ field  since it is absent in our models)
\be
S=\frac{1}{2\p}\int d^2\s\,G_{\m\n}(X)\del_+X^{\m}\del_-X^{\n}.
\end{equation}
The non-vanishing Energy-Momentum tensor components are given by 
\begin{equation}
T_{\pm\pm}=G_{\m\n}\del_{\pm}X^{\m}\del_{\pm}X^{\n}
\end{equation}
satisfying $\del_{\mp}T_{\pm\pm}=0$ on-shell. Due to the previous equations, the components of the Energy-Momentum tensor $T_{\pm\pm}$ satisfy 
\be
\label{Virasoro}
G_{\m\n}\del_{\pm}X^{\m}\del_{\pm}X^{\n}=f_{\pm}(\s_{\pm}),
\end{equation}
These are essentially, first integrals of the equations of motion and, as known, solutions of the equations of motion  also solve \eqref{Virasoro}. In addition, $f_{\pm}(\s_{\pm})$ are functions, which  can be set to constants, denoted as $C_\pm$. Without loss of generality, we may additionally select $C_\pm=C$. Performing a Lorentz boost on the world-sheet we may always restore unequal values of $C_+$ and  $C_-$. The models under study have a positive definite metric, thus these constants are necessarily positive. Notice that in order to embed this model in string theory, one has to consider the tensor product of this theory with another one, so that the overall Energy-Momentum vanishes, implying that the Virasoro constraints are satisfied. Slightly abusing the terminology and having the previous remark in mind, we will refer to \eqref{Virasoro} as the Virasoro constraints.

As mentioned, the Energy-Momentum tensor is conserved on-shell. Nevertheless, the converse is not generally true, unless the target space is 2-dimensional. In this case it follows that
\begin{equation}
\label{linearindep}
\begin{pmatrix}
\partial_-T_{++}\\
\partial_+T_{--}
\end{pmatrix}=\begin{pmatrix}
\partial_+ X^1 & \partial_+ X^2\\
\partial_- X^1 & \partial_- X^2
\end{pmatrix}
\begin{pmatrix}
\frac{\delta\mathcal{L}}{\delta  X^1}\\
\frac{\delta\mathcal{L}}{\delta  X^2}
\end{pmatrix}
,
\end{equation}
where $\mathcal{L}$ is the Lagrangian density and $\d\mathcal{L}/ \d X^{\m}$ is the variation  of the action with respect to the $X^{\m}$ field. Thus, if the matrix on the right-hand-side is invertible, the solutions of the Virasoro constraints are also solutions of the equations of motion \footnote{Of course it is possible that the solutions of the Virassoro constraint still solve the equations of motion, but this is not the case for the solutions of this work.}. This approach resembles the Pohlmeyer reduction, where one gauge fixes the world-sheet diffeomorphisms \cite{Pohlmeyer:1975nb}. 

\subsection{First class of solutions}
Having clarified our methodology, we derive the solutions of the models presented in the previous section.

\subsubsection{$SL(2,\mathbb{R})/U(1)$ model}

\subsubsection*{ Vector Gauging}
\label{sec:sl2_vec_1st}
As a first example we derive classical solutions of the model described by the action \eqref{lWitten_sl2_vector}. The Virasoro constraints \eqref{Virasoro} constitute the pair of differential equations
\be
\label{y1y2}
\frac{1+\l}{1-\l}(\del_{\pm}y_1)^2+\frac{1-\l}{1+\l}(\del_{\pm}y_2)^2=\frac{m^2}{4}(y_1^2+y^2_2-1),
\end{equation}
where we have set $C=m^2/4$ so that the right-hand-side is manifestly positive, as the left-hand-side. The factor of $1/4$ is introduced for future convenience. In order to obtain non-trivial solutions, it is required that $m\neq0$. We can decouple the equations as follows
\begin{align}
&\frac{1+\l}{1-\l}\,(\del_{\pm}y_1)^2=\frac{m^2}{4}\left(y_1^2+c_{\pm}-\frac{1}{2}\right),  \label{y1}\\
& \frac{1-\l}{1+\l}\,(\del_{\pm}y_2)^2=\frac{m^2}{4}\left(y_2^2-c_{\pm}-\frac{1}{2}\right).    \label{y2} 
\end{align}

\begin{table}[htb]
\centering
\begin{tabular}{|c|c|c|}
\hline
 & $y_1(\tau)$ & $y_2(\sigma)$\\
\hline
$c>\frac{1}{2}$ & $\sqrt{c-\frac{1}{2}}\,\,\sinh\left(m\sqrt{\frac{1-\l}{1+\l}}\left(\tau-\tau_0\right)\right)$ & $\sqrt{c+\frac{1}{2}}\,\,\cosh\left(m\sqrt{\frac{1+\l}{1-\l}}\left(\sigma-\sigma_0\right)\right)$\\
\hline
$\frac{1}{2}>c>-\frac{1}{2}$ & $\sqrt{\frac{1}{2}-c}\,\,\cosh\left(m\sqrt{\frac{1-\l}{1+\l}}\left(\tau-\tau_0\right)\right)$ & $\sqrt{c+\frac{1}{2}}\,\,\cosh\left(m\sqrt{\frac{1+\l}{1-\l}}\left(\sigma-\sigma_0\right)\right)$\\
\hline
$-\frac{1}{2}>c$  & $\sqrt{\frac{1}{2}-c}\,\,\cosh\left(m\sqrt{\frac{1-\l}{1+\l}}\left(\tau-\tau_0\right)\right)$ & $\sqrt{-\frac{1}{2}-c}\,\,\sinh\left(m\sqrt{\frac{1+\l}{1-\l}}\left(\sigma-\sigma_0\right)\right)$ \\
\hline
$c=\frac{1}{2}$ & $\mathcal{A}\exp\left(m\sqrt{\frac{1-\l}{1+\l}}\left(\tau-\tau_0\right)\right)$ & $\cosh\left(m\sqrt{\frac{1+\l}{1-\l}}\left(\sigma-\sigma_0\right)\right)$ \\
\hline
$c=-\frac{1}{2}$ & $\cosh\left(m\sqrt{\frac{1-\l}{1+\l}}\left(\tau-\tau_0\right)\right)$ & $\mathcal{A}\exp\left(m\sqrt{\frac{1+\l}{1-\l}}\left(\sigma-\sigma_0\right)\right)$\\
\hline
\end{tabular}
\caption{First class of solution for the  $SL(2,\mathbb{R})/U(1)_V$ $\lambda$-deformed model.}
\label{tab:solutions_sl2_vector}
\end{table}
The equations above have similar form, their only difference is the sign in front of $c_{\pm}$. Thus, solving one equation suffices to obtain solutions for both equations. Considering equation \eqref{y1}, we immediately obtain that
\be
\label{y1eq}
\frac{\del_{\pm}y_1}{\sqrt{y_1^2+c_{\pm}-\frac{1}{2}}}=s_{\pm}\,\frac{m}{2}\sqrt{\frac{1-\l}{1+\l}},
\end{equation}
where $s_{\pm}$ are constants that take independently the value $1$ or $-1$. From now on, we also set $c_{\pm}=c$ \footnote{One can show that this choice is a consistency condition of the equations. In order to keep our expressions as simple a possible, we present the derivation having set the constants $c_+$ and $c_-$ to equal values all along the calculation.}. The functional form of the solution of \eqref{y1eq} depends on the value of the constant $c$. Since is it straightforward to solve equations \eqref{y1eq}, we just present the solutions and make some comments. 

First, the functional form of the solution of \eqref{y1eq} depends on the value of the constant $c$. It turns out that the parametric space is divided in three regions, namely  $c>1/2$, $1/2>c>-1/2$ and $-1/2>c$, along with the limiting values $c=\pm1/2$. Secondly, a quick glimpse on \eqref{y2} suffices in order to realize that the solutions for $y_2$ are obtained from the $y_1$ solutions by sending $c\to -c$ and $\l\to-\l$, while adapting the allowed values for $c$. Finally, one should combine the solutions for $y_1$ and $y_2$, taking into account the requirement that their derivatives are linearly independent as noted in \eqn{linearindep} . Essentially, this condition implies that the solutions for $y_1$ and $y_2$ should not be functions of the same variable. In following we choose $y_1=y_1(\tau)$ and  also discard the various signs $s_\pm$. These can be recovered by parity transformations of the world-sheet coordinates, combined with the discrete symmetries of the action, such as $y_1\rightarrow -y_1$. We gather the solutions appropriately matched in Table \ref{tab:solutions_sl2_vector}. One can verify that these indeed satisfy the equations of motion \eqref{eom1_su2} \footnote{Recall that the $SL(2,\mathbb{R})/U(1)_V$ and the $SU(2)/U(1)$ share the same equations of motion}. Notice that $\tau_0$ and $\s_0$ are integration constants, which may be regarded as collective coordinates of the solutions.

\subsubsection*{Axial Gauging}
\label{sec:sl2_axial_1st}
In the axial gauging case, writing down the Virasoro constraints corresponding  to the action \eqref{lWitten_sl2_axial}, one realizes that the solutions are obtained by setting
\be
c\rightarrow-c,\phantom{000} 1/2\rightarrow -1/2
\end{equation}
on the coefficients of various functions in Table \ref{tab:solutions_sl2_vector}, while neglecting the overall minus sign to compensate the $i$ factor of the transformation \eqref{vector_to_axial}. Note that the range of $c$ should be adjusted appropriately. The solutions are gathered in Table \ref{tab:solutions_sl2_axial}.

\begin{table}[htb]
\centering
\begin{tabular}{|c|c|c|}
\hline
 & $y_1(\tau)$ & $y_2(\sigma)$\\
\hline
$c>\frac{1}{2}$ & $\sqrt{c-\frac{1}{2}}\,\,\cosh\left(m\sqrt{\frac{1-\l}{1+\l}}\left(\tau-\tau_0\right)\right)$ & $\sqrt{c+\frac{1}{2}}\,\,\sinh\left(m\sqrt{\frac{1+\l}{1-\l}}\left(\sigma-\sigma_0\right)\right)$\\
\hline
$\frac{1}{2}>c>-\frac{1}{2}$ & $\sqrt{\frac{1}{2}-c}\,\,\sinh\left(m\sqrt{\frac{1-\l}{1+\l}}\left(\tau-\tau_0\right)\right)$ & $\sqrt{c+\frac{1}{2}}\,\,\sinh\left(m\sqrt{\frac{1+\l}{1-\l}}\left(\sigma-\sigma_0\right)\right)$\\
\hline
$-\frac{1}{2}>c$  & $\sqrt{\frac{1}{2}-c}\,\,\sinh\left(m\sqrt{\frac{1-\l}{1+\l}}\left(\tau-\tau_0\right)\right)$ & $\sqrt{-\frac{1}{2}-c}\,\,\cosh\left(m\sqrt{\frac{1+\l}{1-\l}}\left(\sigma-\sigma_0\right)\right)$ \\
\hline
$c=\frac{1}{2}$ & $\mathcal{A}\exp\left(m\sqrt{\frac{1-\l}{1+\l}}\left(\tau-\tau_0\right)\right)$ & $\sinh\left(m\sqrt{\frac{1+\l}{1-\l}}\left(\sigma-\sigma_0\right)\right)$ \\
\hline
$c=-\frac{1}{2}$ & $\sinh\left(m\sqrt{\frac{1-\l}{1+\l}}\left(\tau-\tau_0\right)\right)$ & $\mathcal{A}\exp\left(m\sqrt{\frac{1+\l}{1-\l}}\left(\sigma-\sigma_0\right)\right)$\\
\hline
\end{tabular}
\caption{First class of solution for the  $SL(2,\mathbb{R})/U(1)_A$ $\lambda$-deformed model.}
\label{tab:solutions_sl2_axial}
\end{table}

\subsubsection{ $SU(2)/U(1)$ model}
\label{sec:su2_1st}
The solutions of the $SU(2)/U(1)$ model are obtain by the ones of $SL(2,\mathbb{R})/U(1)_V$ case, for $m^2\to -m^2$. As this coset is compact, the admissible values of the parameter $c$, are bounded. In particular it turns out that $1/2\geq c\geq-1/2$. The solution are gathered in Table \ref{tab:solutions_su2}. Notice that all solutions in Table \ref{tab:solutions_sl2_vector} and \ref{tab:solutions_sl2_axial} (except the exponential ones) can be cast in the form given above.

\begin{table}[htb]
\centering
\begin{tabular}{|c|c|c|}
\hline
 & $y_1(\tau)$ & $y_2(\sigma)$\\
\hline
$\frac{1}{2}>c>-\frac{1}{2}$ & $\sqrt{\frac{1}{2}-c}\,\,\cos\left(m\sqrt{\frac{1-\l}{1+\l}}\,\,\left(\tau-\tau_0\right)\right)$ & $\sqrt{\frac{1}{2}+c}\,\,\cos\left(m\sqrt{\frac{1+\l}{1-\l}}\,\,\left(\sigma-\sigma_0\right)\right)$\\
\hline
\end{tabular}
\caption{First class of solution for the $SU(2)/U(1)$ $\lambda$-deformed model.}
\label{tab:solutions_su2}
\end{table}

\subsection{Second class of solutions}
The second class of solutions is expressed in terms of Jacobi elliptic functions. A short review is given in Appendix \ref{sec:Jacobi_review}
where all the necessary definitions are provided, along with several of their properties. A careful study will help the unfamiliar reader  for the better understanding of the material that follows.

\subsubsection{$SL(2,\mathbb{R})/U(1)$ model}

\subsubsection*{ Vector Gauging}
\label{sec:sl2_vec_2nd}

Let us now derive solutions for the action \eqref{eq:action_sl2_chi_psi_vector}. We implement the same strategy that we used for the first class of solutions by solving the Virasoro constraints and combining the solutions appropriately to satisfy the equations of motion. Again, the form of the metric implies that the constants appearing at the Virasoro constraint have to be positive definite. Thus, in the following we obtain solutions of the equations
\be
\label{quadchipsi}
(\del_{\pm}\chi)^2+(\del_{\pm}\psi)^2=\frac{m^2}{4}\left(\frac{1+\l}{1-\l}-\frac{4\l}{1-\l^2}\cos^2\psi-e^{-2\chi}\right).
\end{equation}
Equations \eqref{quadchipsi} are decoupled as
\begin{align}
&(\del_{\pm}\chi)^2+\frac{m^2}{4}e^{-2\chi}=\frac{m^2}{4}c^2_{\pm},\label{eq:chi_eom_sl2}\\
&(\del_{\pm}\psi)^2-\frac{m^2}{4}\left(\frac{1+\l}{1-\l}-\frac{4\l}{1-\l^2}\cos^2\psi\right)=-\frac{m^2}{4}c^2_{\pm}.\label{eq:psi_eom_sl2}
\end{align}
Positivity of the  constants on the right-hand-side of \eqref{eq:chi_eom_sl2} is required, due to the manifest positivity of the left-hand-side. Immediately, it follows that $\chi$ is given by
\be
e^{\chi}=\frac{1}{c}\cosh\left[m\,c\left(\frac{\s^+\pm \s^-}{2}+\a\right)\right],
\end{equation}
where $\a$ is an integration constant. Notice that the equations \eqref{eq:chi_eom_sl2} are incompatible, unless $c_+=c_-=c>0$. Moving to the second pair of equations we can factorise it as
\begin{align}
&(\del_{\pm}\psi)^2=\frac{m^2}{4}\ell^2\left(1-\kappa^2\cos^2\psi\right), \quad \lambda>0,\\
&(\del_{\pm}\psi)^2=\frac{m^2}{4}\ell^2\left(1-\kappa^2\sin^2\psi\right), \quad \lambda<0,
\end{align}
where $\kappa^2$ is the elliptic modulus, which is given by 
\begin{equation}\label{eq:elliptic_modulus_sl2}
\kappa^2=\frac{4\vert\l\vert}{1-\l^2}\,\frac{1}{\ell^2}
\end{equation}
and $\ell$ is defined by the equation
\begin{equation}\label{eq:ell_sl2v}
\ell^2=\frac{1+\vert\l\vert}{1-\vert\l\vert}-c^2.
\end{equation}
As these equations are of the form of \eqref{eq:Jacobi_ode}, it follows that their solution is
\begin{equation}\label{eq:solution_psi_am_vec}
\psi=\mathrm{am}\left[m\,\ell\left(\frac{\sigma^+\mp\sigma^-}{2}+\beta\right)\Big\vert \kappa^2\right]+\frac{\pi}{2}\frac{\lambda+\vert\lambda\vert}{2\vert\lambda\vert},
\end{equation}
where $\mathrm{am}(x\vert m)$ is the Jacobi amplitude\footnote{See Appendix \ref{sec:Jacobi_review} for a review about basic properties of the Jacobi elliptic functions.}. The reality of the solutions requires $\ell^2\geq0$, which implies that $c$ is subject to the constraint
\begin{equation}\label{eq:c_constraint_vec}
\sqrt{\frac{1+\vert \lambda\vert}{1-\vert\lambda\vert}}\geq c\geq0.
\end{equation}
Choosing $\chi=\chi(\tau)$ (and necessarily $\psi=\psi(\s)$) we obtain the solution
\begin{align}
e^{\chi(\tau)}&=\frac{1}{c}\cosh\left[m\,c\left(\tau-\tau_0\right)\right],\label{eq:exp_chi}\\
\psi(\sigma)&=\mathrm{am}\left[m\,\ell\left(\sigma-\sigma_0\right)\Big\vert \kappa^2\right]+\frac{\pi}{2}\frac{\lambda+\vert\lambda\vert}{2\vert\lambda\vert}\label{eq:psi_jacobi}.
\end{align}
Of course, the class of solutions $\chi=\chi(\sigma)$ and $\psi=\psi(\tau)$ is also admissible.

\subsubsection*{Axial gauging}
\label{sec:sl2_axial_2nd}
For the axial gauging  we have to  solve the Virasoro constraints
\be
\label{quadchipsi_ax}
(\del_{\pm}\chi)^2+(\del_{\pm}\psi)^2=\frac{m^2}{4}\left(\frac{1+\l}{1-\l}-\frac{4\l}{1-\l^2}\cos^2\psi+e^{-2\chi}\right).
\end{equation}
Notice that both contributions of the right-hand-side are manifestly positive. This will have interesting consequences at the set of solutions. Decoupling the equations as in the case of vector gauging, there are two distinct pairs of equations
\begin{align}
&(\del_{\pm}{\chi})^2-\frac{m^2}{4}e^{-2{\chi}}=\pm\frac{m^2}{4}c^2_{\pm},\label{eq:chi_eom_sl2_ax1},\\
&(\del_{\pm}\psi)^2-\frac{m^2}{4}\left(\frac{1+\l}{1-\l}-\frac{4\l}{1-\l^2}\cos^2\psi\right)=\mp\frac{m^2}{4}c^2_{\pm},\label{eq:psi_eom_sl2_ax1},
\end{align}
since the left-hand-side of \eqref{eq:chi_eom_sl2_ax1} is not manifestly positive. Again, the equations are incompatible unless $c_+=c_-=c$. Choosing first, the plus sign in \eqref{eq:chi_eom_sl2_ax1} we obtain the solution
\begin{equation}
e^{{\chi}}=\frac{1}{c}\sinh\left[m\,c\left(\frac{\s^+\pm \s^-}{2}+\a\right)\right].
\end{equation}
The solution for $\psi$ is provided by \eqref{eq:solution_psi_am_vec}, where $c$ is subjected to equation \eqref{eq:c_constraint_vec}. Of course, $c$ may be taken purely imaginary to provide a solution for the minus sign in \eqref{eq:chi_eom_sl2_ax1}. In such a case the solution is automatically real and the parameter $c$ is unconstrained. Putting everything together, we obtain the following class of solutions
\begin{align}
e^{{\chi}_1(\tau)}&=\frac{1}{c}\sinh\left[m\,c\left(\tau-\tau_0\right)\right],\label{eq:exp_chi_ax1}\\
\psi_1(\sigma)&=\mathrm{am}\left[m\,\ell\left(\sigma-\sigma_0\right)\Big\vert \kappa^2\right]+\frac{\pi}{2}\frac{\lambda+\vert\lambda\vert}{2\vert\lambda\vert}\label{eq:psi_jacobi_ax1},
\end{align}
where the elliptic modulus $\kappa^2$ and $\ell$ are defined in \eqref{eq:elliptic_modulus_sl2} and \eqref{eq:ell_sl2v}, respectively, as well as,
\begin{align}
e^{\chi_2(\tau)}&=\frac{1}{c}\sin\left[m\,c\left(\tau-\tau_0\right)\right],\label{eq:exp_chi_ax2}\\
\psi_2(\sigma)&=\mathrm{am}\left[m\,\tilde{\ell}\left(\sigma-\sigma_0\right)\Big\vert \tilde{\kappa}^2\right]+\frac{\pi}{2}\frac{\lambda+\vert\lambda\vert}{2\vert\lambda\vert}\label{eq:psi_jacobi_ax2},
\end{align}
where the elliptic modulus $\tilde{\kappa}^2$ and $\tilde{\ell}$ are defined as
\begin{equation}\label{eq:elliptic_modulus_tilde}
\tilde{\kappa}^2=\kappa^2\vert_{\ell^2\rightarrow \tilde\ell^2}\,\,, \qquad \tilde{\ell}^2=\ell^2\vert_{c^2\rightarrow -c^2}\,\,.
\end{equation}
Notice that this solution is valid for any $c$. Of course, one can interchange $\sigma$ and $\tau$ to obtain the rest of the solutions belonging to these classes.

\subsubsection{ $SU(2)/U(1)$ model}
\label{sec:su2_2nd}
As a final step we proceed with the solutions on the $SU(2)/U(1)$. Proceeding in the usual manner, by solving the Virasoro constraints, the decoupled equations read
\begin{align}
&(\del_{\pm}\chi)^2-\frac{m^2}{4}e^{-2\chi}=-\frac{m^2}{4}c^2_{\pm},\label{eq:chi_eom_su2}\\
&(\del_{\pm}\psi)^2+\frac{m^2}{4}\left(\frac{1+\l}{1-\l}-\frac{4\l}{1-\l^2}\cos^2\psi\right)=\frac{m^2}{4}c^2_{\pm},\label{eq:psi_eom_su2}
\end{align}
where the right-hand-side of \eqref{eq:psi_eom_su2} is manifestly positive in view of \eqref{eq:cos_inequality}. The solution of the first equation is
\begin{equation}
e^{\chi}=\frac{1}{c}\sin\left[m\,c\left(\frac{\s^+\pm \s^-}{2}+a\right)\right],
\end{equation}
while the pair of equations \eqref{eq:psi_eom_su2} can be written as
\begin{align}
(\del_{\pm}\psi)^2=\frac{m^2}{4}\bar{\ell}^2\left(1-\bar{\kappa}^2\sin^2\psi\right), \quad \lambda>0,\\
(\del_{\pm}\psi)^2=\frac{m^2}{4}\bar{\ell}^2\left(1-\bar{\kappa}^2\cos^2\psi\right), \quad \lambda<0,
\end{align}
where the elliptic modulus $\bar{\kappa}^2$ is defined as
\begin{equation}\label{eq:elliptic_modulus_bar}
\bar{\kappa}^2=\frac{4\vert\l\vert}{1-\l^2}\,\frac{1}{\bar\ell^2}
\end{equation}
and $\bar{\ell}$ is defined via the equation
\begin{equation}
\bar{\ell}^2=c^2-\frac{1-\vert\l\vert}{1+\vert\l\vert}.
\end{equation} 
The reality of the solutions implies that $c$ is subject to the constraint
\begin{equation}\label{eq:c_constraint_su2}
c^2\geq \frac{1-\vert\l\vert}{1+\vert\l\vert}.
\end{equation}
Considering $\chi=\chi(\tau)$ and $\psi=\psi(\sigma)$, the solution reads
\begin{align}
e^{\chi(\tau)}&=\frac{1}{c}\sin\left[m\,c\left(\tau-\tau_0\right)\right],\label{eq:solution_chi_su2}\\
\psi(\sigma)&=\mathrm{am}\left[m\,\bar{\ell}\left(\sigma-\sigma_0\right)\Big\vert \bar{\kappa}^2\right]+\frac{\pi}{2}\frac{\vert\lambda\vert-\lambda}{2\vert\lambda\vert}\,\,\,.\label{eq:solution_psi_am_su2}
\end{align}

\section{Properties of the Solutions}
\label{sec:properties}
In this section we study the properties of the solutions we derived in the previous one. First of all, we summarize the second class of solutions and present their basic features. Then, we specify the boundary conditions. This is required in order to eliminate the surface terms, rising when varying the action. In addition, since $\sigma$-models describe string theories\footnote{In order to do so one has to tensor the models of this work, with another sigma model corresponding to a metric of indefinite signature. The simplest case is to consider a single time dimension. In this case, the equations of motion are solved by $t=\frac{1}{2}\left(m_+\sigma_+ +m_-\sigma_-\right)$. This selection also makes the overall Virasoro constraints vanish, implying that the theory is (classically) conformally invariant. Had we considered $C_+\neq C_-$ in \eqref{Virasoro}, the obtained solutions would depend on this coordinate and on $\Sigma=\frac{1}{2}\left(m_+\sigma_+ -m_-\sigma_-\right)$. Thus, from the target space perspective, the values of of $m_+$ and $m_-$ are irrelevant, whereas on the world-sheet their values may be altered by a Lorentz boost.}, we also mention possible brane configurations, related to the aforementioned boundary conditions. Having specified all admissible cases, we plot the solutions and discuss them. Finally,  the effect of the non-perturbative dualities \eqref{duality} and \eqref{-lsymmetry} is described. 

\begin{table}[htb]
\centering
\begin{tabular}{|c|c|c|c|c|}
\hline
Model & $\chi(\tau)$ & $\psi(\sigma)$ & Elliptic Modulus & Constraint on $c$\\
\hline
$SL(2,\mathbb{R})/U(1)_V$ & \eqref{eq:exp_chi} & \eqref{eq:psi_jacobi} & \eqref{eq:elliptic_modulus_sl2} & \eqref{eq:c_constraint_vec}\\
\hline
$SL(2,\mathbb{R})/U(1)_{A_1}$ & \eqref{eq:exp_chi_ax1} & \eqref{eq:psi_jacobi_ax1} & \eqref{eq:elliptic_modulus_sl2} & \eqref{eq:c_constraint_vec}\\
\hline
$SL(2,\mathbb{R})/U(1)_{A_2}$ & \eqref{eq:exp_chi_ax2} & \eqref{eq:psi_jacobi_ax2} & \eqref{eq:elliptic_modulus_tilde} & -\\
\hline
$SU(2)/U(1)$ & \eqref{eq:solution_chi_su2} & \eqref{eq:solution_psi_am_su2} & \eqref{eq:elliptic_modulus_bar} & \eqref{eq:c_constraint_su2}\\
\hline
\end{tabular}
\caption{This table summarizes the expressions for the coordinates $\chi$ and $\psi$, and the elliptic modulus for each of the models of the first column. Notice that there are two kinds of solutions for the $SL(2,\mathbb{R})/U(1)$ axially gauged model. There is no constraint on $c$ for the second kind of solutions.}
\label{tab:solutions_2nd}
\end{table}

\begin{table}[htb]
\centering
\begin{tabular}{|c|c|c|c|c|c|c|}
\hline
Model & $\chi$ & $\psi$ &  & Phase & Range of $c$ & Plot\\
\hline
$SL(2,\mathbb{R})/U(1)_V$ & $\chi(\tau)$ & $\psi(\sigma)$ & ST & Oscillating & $\frac{1-\vert\lambda\vert}{1+\vert\lambda\vert} \leq c^2\leq \frac{1+\vert\lambda\vert}{1-\vert\lambda\vert}$ & Fig. \ref{fig:SL2V_static}\\
\hline
$SL(2,\mathbb{R})/U(1)_V$ & $\chi(\tau)$ & $\psi(\sigma)$ & ST & Rotating & $0 \leq c^2\leq \frac{1-\vert\lambda\vert}{1+\vert\lambda\vert}$ & Fig. \ref{fig:SL2V_static}\\
\hline
$SL(2,\mathbb{R})/U(1)_V$ & $\chi(\sigma)$ & $\psi(\tau)$ & TI & Oscillating & $\frac{1-\vert\lambda\vert}{1+\vert\lambda\vert} \leq c^2\leq \frac{1+\vert\lambda\vert}{1-\vert\lambda\vert}$ & Fig. \ref{fig:SL2V_ti}\\
\hline
$SL(2,\mathbb{R})/U(1)_V$ & $\chi(\sigma)$ & $\psi(\tau)$ & TI & Rotating & $0 \leq c^2\leq \frac{1-\vert\lambda\vert}{1+\vert\lambda\vert}$& Fig. \ref{fig:SL2V_ti}\\
\hline
$SL(2,\mathbb{R})/U(1)_{A_1}$ & $\chi(\tau)$ & $\psi(\sigma)$ & ST & Oscillating & $\frac{1-\vert\lambda\vert}{1+\vert\lambda\vert} \leq c^2\leq \frac{1+\vert\lambda\vert}{1-\vert\lambda\vert}$& Fig. \ref{fig:SL2A_static}\\
\hline
$SL(2,\mathbb{R})/U(1)_{A_1}$ & $\chi(\tau)$ & $\psi(\sigma)$ & ST & Rotating & $0 \leq c^2\leq \frac{1-\vert\lambda\vert}{1+\vert\lambda\vert}$& Fig. \ref{fig:SL2A_static}\\
\hline
$SL(2,\mathbb{R})/U(1)_{A_1}$ & $\chi(\sigma)$ & $\psi(\tau)$ & TI & Oscillating & $\frac{1-\vert\lambda\vert}{1+\vert\lambda\vert} \leq c^2\leq \frac{1+\vert\lambda\vert}{1-\vert\lambda\vert}$& Fig. \ref{fig:SL2A_ti}\\
\hline
$SL(2,\mathbb{R})/U(1)_{A_1}$ & $\chi(\sigma)$ & $\psi(\tau)$ & TI & Rotating & $0 \leq c^2\leq \frac{1-\vert\lambda\vert}{1+\vert\lambda\vert}$& Fig. \ref{fig:SL2A_ti}\\
\hline
$SL(2,\mathbb{R})/U(1)_{A_2}$ & $\chi(\tau)$ & $\psi(\sigma)$ & ST & Rotating & $c\in\mathbb{R}$ & Fig. \ref{fig:SL2A2}\\
\hline
$SL(2,\mathbb{R})/U(1)_{A_2}$ & $\chi(\sigma)$ & $\psi(\tau)$ & TI & Rotating & $c\in\mathbb{R}$ & Fig. \ref{fig:SL2A2}\\
\hline
$SU(2)/U(1)$ & $\chi(\tau)$ & $\psi(\sigma)$ & ST & Oscillating & $\frac{1-\vert\lambda\vert}{1+\vert\lambda\vert}\leq c^2\leq\frac{1+\vert\lambda\vert}{1-\vert\lambda\vert}$ & Fig. \ref{fig:SU2_static}\\
\hline
$SU(2)/U(1)$ & $\chi(\tau)$ & $\psi(\sigma)$ & ST & Rotating & $\frac{1+\vert\lambda\vert}{1-\vert\lambda\vert}\leq c^2$& Fig. \ref{fig:SU2_static}\\
\hline
$SU(2)/U(1)$ & $\chi(\sigma)$ & $\psi(\tau)$ & TI & Oscillating & $\frac{1-\vert\lambda\vert}{1+\vert\lambda\vert}\leq c^2\leq\frac{1+\vert\lambda\vert}{1-\vert\lambda\vert}$& Fig. \ref{fig:SU2_ti}\\
\hline
$SU(2)/U(1)$ & $\chi(\sigma)$ & $\psi(\tau)$ & TI & Rotating & $\frac{1+\vert\lambda\vert}{1-\vert\lambda\vert}\leq c^2$& Fig. \ref{fig:SU2_ti}\\
\hline
\end{tabular}
\caption{All solutions  of the second class are classified by their characteristics. Solutions are characterized either as static (ST) if $\psi=\psi(\sigma)$ or as translationally invariant (TI) if  $\psi=\psi(\tau)$. Similarly, they are characterized as oscillating if the value of the elliptic modulus is greater than 1 and as rotating if it is between zero and one. The expressions for the coordinates $\chi$ and $\psi$, and the elliptic modulus for each of the models of the first column are in Table \ref{tab:solutions_2nd}.}
\label{tab:solutions_2nd_categories}
\end{table}

\subsection{Overview of the 2nd Class of Solutions}
\label{sec:2nd_class_summary}
In this section we summarize the second class of solutions and discuss some of their common features. Before doing so, we present the coordinates $y_1$ and $y_2$ for this class of solutions. Taking into account the definitions of $y_1$ and $y_2$ for each model, namely equations \eqref{apcoor_su2}, \eqref{apcoor_sl2_vector} and \eqref{apcoor_sl2_axial}, along with \eqref{eq:psi_to_phi}, which is common for all models, as well as the equations \eqref{eq:r_vec}, \eqref{eq:r_axial} and \eqref{eq:theta}, it follows that $y_1$ and $y_2$ are given in terms of $\chi$ and $\psi$ by
\begin{equation}
y_1=\sqrt{\frac{1-\lambda}{1+\lambda}}e^{\chi}\cos\psi,\qquad y_2=\sqrt{\frac{1+\lambda}{1-\lambda}}e^{\chi}\sin\psi,
\end{equation}
for all models. This expressions are free of the subtleties regarding  the positivity of $e^\chi$, which appear in equations  \eqref{eq:exp_chi_ax1}, \eqref{eq:exp_chi_ax2} and \eqref{eq:solution_chi_su2}. It is evident that this class of solutions parametrize an ellipsis. The eccentricity of this ellipsis depends on $\lambda$ \footnote{Nevertheless the effect of the $\lambda-$ deformation is much more than a rescaling of coordinates, as the elliptic modulus depends non-trivially on it.}. In the case $\chi=\chi(\tau)$ the overall scale of the ellipsis is time dependent, whereas in the case $\chi=\chi(\sigma)$, the solution is a rod, whose endpoint(s) lies on ellipses. In both cases, it is straightforward to show that these solutions satisfy
\begin{equation}
G_{y_1 y_1}\partial_{\sigma}y_1\partial_{\tau}y_1+G_{y_2 y_2}\partial_{\sigma}y_2\partial_{\tau}y_2=0,
\end{equation}
for all $\s,\tau$. This implies that when considering open strings, they are always perpendicular to the surfaces  traced by their endpoints. This equation is a direct consequence of the Virasoro constraints. The Virasoro constraints also imply
\begin{equation}
G_{y_1 y_1}\left[\left(\partial_{\tau}y_1\right)^2+\left(\partial_{\sigma}y_1\right)^2\right]+G_{y_2 y_2}\left[\left(\partial_{\tau}y_2\right)^2+\left(\partial_{\sigma}y_2\right)^2\right]=m^2,\quad \forall \sigma,\tau,
\end{equation}
the physical time being $t=m\tau$. Thus, there is no momentum flow on strings endpoints. The exact behaviour of the solution depends on the values of the parameters and the boundary conditions. In Table \ref{tab:solutions_2nd} we gather all equations, which define the second class of solutions for each model. Table  \ref{tab:solutions_2nd_categories}   presents the classification of all solutions of the second class according to their features. These include the cases $\psi=\psi(\sigma)$ and  $\psi=\psi(\tau)$, as well as whether the value of the elliptic modulus is greater or smaller that one. This classification is analogous to the one performed in \cite{Katsinis:2018zxi} regarding elliptic string solutions in $\mathbb{R}\times\textrm{S}^2$.

For the solutions of  $SL(2,\mathbb{R})/U(1)_V$ model and one of the two solutions of the $SL(2,\mathbb{R})/U(1)_A$ model, the elliptic modulus $\kappa^2$, defined in \eqref{eq:elliptic_modulus_sl2}, satisfies $0\leq\kappa^2\leq 1$ when
\begin{equation}\label{eq:ksq_range_rot}
\frac{1-\vert\lambda\vert}{1+\vert\lambda\vert} \leq c^2\leq \frac{1+\vert\lambda\vert}{1-\vert\lambda\vert}
\end{equation}
and is associated to rotating solutions. Similarly, when
\begin{equation}\label{eq:ksq_range_osc}
0 \leq c^2\leq \frac{1-\vert\lambda\vert}{1+\vert\lambda\vert}
\end{equation}
it satisfies $1\leq \kappa^2$ and it is related to oscillating solutions. For these solutions we define
\begin{equation}\label{eq:ds_definition}
\delta\sigma=\frac{\omega_1}{m\ell}\,\,\,.
\end{equation}
The half-period $\omega_1$ is defined in terms of the elliptic modulus via \eqref{eq:w1_definition}.

Considering the other solution of the  $SL(2,\mathbb{R})/U(1)_A$ model, the elliptic modulus $\tilde{\kappa}^2$, defined in \eqref{eq:elliptic_modulus_tilde}, always satisfies $0\leq\tilde{\kappa}^2\leq1$ for any value of $c$. In this case we define
\begin{equation}\label{eq:ds_tilde_definition}
\delta\tilde{\sigma}=\frac{\omega_1}{m\tilde\ell}\,\,.
\end{equation}
Finally, for the solutions of the $SU(2)/U(1)$ model, the elliptic modulus $\bar{\kappa}^2$, defined in \eqref{eq:elliptic_modulus_bar}, satisfies $0\leq\bar{\kappa}^2\leq 1$ when
\begin{equation}\label{eq:ksq_tilde_range_rot}
\frac{1+\vert\lambda\vert}{1-\vert\lambda\vert}\leq c^2
\end{equation}
related to rotating solutions. Similarly, it satisfies $1\leq \bar{\kappa}^2$ when
\begin{equation}\label{eq:ksq_tilde_range_osc}
\frac{1-\vert\lambda\vert}{1+\vert\lambda\vert} \leq c^2\leq \frac{1+\vert\lambda\vert}{1-\vert\lambda\vert}
\end{equation}
and is associated to oscillating solutions. Finally, in this case we define
\begin{equation}\label{eq:ds_bar_definition}
\delta\bar{\sigma}=\frac{\omega_1}{m\bar\ell}\,\,.
\end{equation}
The lengths $\delta\sigma$, $\delta\tilde{\sigma}$ and $\delta\bar{\sigma}$ will be used when studying the boundary conditions and the periodicity properties.

\subsection*{Special Limits}
Solutions of the second class have two interesting limits. The first one is the limit of the vanishing elliptic modulus. This limit is obtained for $\lambda=0$. In this case the Jacobi amplitude becomes just the linear function $x$, see equation \eqref{eq:vacuum_limit}. The elliptic function degenerate to trigonometric ones. In general, elliptic functions are defined on a torus, since they are doubly periodic. In this limit the imaginary period diverges and the torus becomes singular. Starting from the undeformed solution, from a mathematical point of view, the $\lambda-$deformation resolves this singularity and the corresponding degenerate torus becomes non-degenerate.

A far more interesting limit is the one of the diverging real period. This is the case when the elliptic modulus equals to unity. The Jacobi amplitude is given by \eqref{eq:kink_limit}. The elliptic functions degenerate to hyperbolic ones. The form of the solutions in this limit is presented in Table \ref{tab:solutions_2nd_kink}. Notice that we present only the $\lambda>0$ solutions, while the rest of them are obtained using the duality \eqref{-lsymmetry} ($m$ is invariant). We also present only the static solutions. The translationally invariant ones are obtained via the $\sigma\leftrightarrow\tau$ transformation.

\begin{table}[htb]
\centering
\begin{tabular}{| c | c | c |}
\hline
Model & $y_1$ & $y_2$\\
\hline
$SL(2,\mathbb{R})/U(1)_V$ & $\frac{\cosh\left(m\sqrt{\frac{1-\lambda}{1+\lambda}}\tau\right)}{\coth\left(2m\sqrt{\frac{\lambda}{1-\lambda^2}}\sigma\right)}$ & $\frac{1+\lambda}{1-\lambda}\frac{\cosh\left(m\sqrt{\frac{1-\lambda}{1+\lambda}}\tau\right)}{\cosh\left(2m\sqrt{\frac{\lambda}{1-\lambda^2}}\sigma\right)}$\\
\hline
$SL(2,\mathbb{R})/U(1)_A$ & $\frac{\sinh\left(m\sqrt{\frac{1-\lambda}{1+\lambda}}\tau\right)}{\coth\left(2m\sqrt{\frac{\lambda}{1-\lambda^2}}\sigma\right)}$ & $\frac{1+\lambda}{1-\lambda}\frac{\sinh\left(m\sqrt{\frac{1-\lambda}{1+\lambda}}\tau\right)}{\cosh\left(2m\sqrt{\frac{\lambda}{1-\lambda^2}}\sigma\right)}$\\
\hline
$SU(2)/U(1)$ & $\frac{1-\lambda}{1+\lambda}\frac{\sin\left(m\sqrt{\frac{1+\lambda}{1-\lambda}}\tau\right)}{\cosh\left(2m\sqrt{\frac{\lambda}{1-\lambda^2}}\sigma\right)}$ & $\frac{\sin\left(m\sqrt{\frac{1+\lambda}{1-\lambda}}\tau\right)}{\coth\left(2m\sqrt{\frac{\lambda}{1-\lambda^2}}\sigma\right)}$\\
\hline
\end{tabular}
\caption{The kink limit of the static solutions of the second class for $\lambda>0$. Translationally invariant ones are obtained via $\sigma\leftrightarrow\tau$, while the $\lambda<0$, are obtained using the duality \eqref{-lsymmetry}.}
\label{tab:solutions_2nd_kink}
\end{table}

\subsection{Boundary conditions - Analysis: 1st class}
Up to now we have not addressed the problem of surface terms of the actions. Besides the validity of the solution per se, considering the embedding of the sigma model in string theory, these boundary conditions may describe either open or closed strings and hint at the brane configurations, which are associated with the string solution. Note that our target space is two-dimensional restricting us in p2-branes as boundary configurations at most. The boundary terms, dropped, when varying the actions \eqref{lWitten_su2}, \eqref{lWitten_sl2_vector} and \eqref{lWitten_sl2_axial}, are
\begin{equation}
\label{boundaryc}
\delta y_i e^{2\Phi}\partial_\sigma y_i\big\vert_{\sigma=\sigma_i}=\delta y_i e^{2\Phi}\partial_\sigma y_i\big\vert_{\sigma=\sigma_f},
\end{equation}
where $i=1,2$. 

We first present the boundary conditions of the first class and then discuss the solutions of the second class. As the expressions for the first class are very simple, we can treat both vector and axial gauging simultaneously. Regarding the second class of solutions we use the classification introduced in Section \ref{sec:2nd_class_summary}. The reader may find Tables \ref{tab:solutions_2nd} and \ref{tab:solutions_2nd_categories} particularly useful, since they provide an overview of the solutions, their classification and corresponding range of the parameters. As a last remark, for convenience we consider that for the second class of solutions the parameters $\sigma_0$ and $\tau_0$ vanish.

\subsubsection*{$SL(2,\mathbb{R})/U(1)$ models }
The solutions we obtained, either in the case of vector gauging (Table \ref{tab:solutions_sl2_vector}) or in the case of axial gauging (Table \ref{tab:solutions_sl2_axial}), are hyperbolic and exponentials functions\footnote{The first subscript  denotes the type of function and the second one the coordinate of the function.} 
\begin{equation}
y_{s,\tau}=a\sinh\left(b\left(\tau-\tau_0\right)\right),\quad y_{c,\tau}=a\cosh\left(b\left(\tau-\tau_0\right)\right),\quad y_{e,\tau}=a\exp\left(b\left(\tau-\tau_0\right)\right)
\end{equation}
and
\begin{equation}
y_{s,\sigma}=a^\prime\sinh\left(b^\prime\left(\sigma-\sigma_0\right)\right),\quad y_{c,\sigma}=a^\prime\cosh\left(b^\prime\left(\sigma-\sigma_0\right)\right),\quad y_{e,\sigma}=a^\prime\exp\left(b^\prime\left(\sigma-\sigma_0\right)\right).
\end{equation}
satisfying  $e^{2\Phi}\partial_\sigma y_{i,\alpha}\big\vert_{\sigma=\pm\infty}=0$, where $i=c,s,e$ and $\alpha=\sigma,\tau$. Thus, the variational problem is well defined for $\sigma\,\in \,(-\infty,\infty)$ without imposing any further conditions and these solutions naturally represent long strings. The world-sheet of such solutions is the Minkowski plane.

One can impose boundary conditions in the following cases:
\paragraph{Neumann:}  
Since 
\begin{equation}
\partial_\sigma y_{c,\sigma}\big\vert_{\sigma=\sigma_0}=0,
\end{equation}
we can impose Neumann conditions both for $y_{c,\sigma}$ and $y_{I,\tau}$ at $\sigma=\sigma_0$.

\paragraph{Dirichlet:}  
Since 
\begin{equation}
\partial_\tau y_{I,\sigma}=0,\qquad \partial_\sigma y_{I,\tau}=0,
\end{equation}
we can impose Dirichlet conditions for $y_{I,\sigma}$ and Neumann for $y_{I^\prime,\tau}$ for any arbitrary $\sigma$.

Combining appropriately the above two cases, we can impose the following boundary conditions on the solutions:
\begin{itemize}
\item N for $y_{c,\sigma}$ and $y_{I,\tau}$ at $\sigma=\sigma_0$, corresponding to a space filling p2-brane.
\item D for $y_{I,\sigma}$ and N for $y_{I^\prime,\tau}$ at $\sigma=\sigma_D$, corresponding to semi-infinite sting ending on a single p1-brane.
\item D-D for $y_{I,\sigma}$ and N-N $y_{I^\prime,\tau}$ at $\sigma=\sigma_i$ and $\sigma=\sigma_f$, corresponding to a pair of p1-branes.
\item D-N for $y_{c,\sigma}$ and N-N $y_{I,\tau}$ at $\sigma=\sigma_D$ and $\sigma=\sigma_0$, corresponding to a p1-brane and a space filling p2-brane.
\end{itemize}
All these solutions correspond to infinite, semi-infinite or finite moving line segments. Intestingly enough, the D-D and N-N boundary conditions are integrable \cite{Driezen:2019ykp}.

\subsubsection*{$SU(2)/U(1)$ model}

Contrary to the $SL(2;\mathbb{R})/U(1)$ case, the conformal factor $e^{2\Phi}$, defined in \eqref{lWitten_su2}, does not vanish at $\sigma=\pm\infty$. This class of solutions is naturally periodic, corresponding to closed configurations. We discuss, the $y_1=y_1(\tau)$ and $y_2=y_2(\sigma)$ case, but similar conclusions hold for the other case too. In particular, the solutions (Table \ref{tab:solutions_su2}) are periodic under
\begin{equation}
\sigma\rightarrow\sigma+\delta\sigma,\qquad \delta\sigma=\frac{2\pi}{m}\sqrt{\frac{1-\l}{1+\l}}.
\end{equation}
These configurations are folded strings, which as time flows, oscillate in the $y_1$ direction. The world-sheet of such solutions is a torus.

One can impose open string boundary conditions in the following cases:
\paragraph{Neumann:}  
Since 
\begin{equation}
\partial_\sigma y_2\big\vert_{\sigma=\sigma_N}=0,\qquad \sigma_N=\sigma_0+n\sqrt{\frac{1-\l}{1+\l}}\frac{\pi}{m},\qquad n\in\mathbb{N}
\end{equation}
we can impose Neumann conditions both for $y_1$ and $y_2$ at $\sigma=\sigma_N$.

\paragraph{Dirichlet:}  
Since 
\begin{equation}
\partial_\tau y_2=0,\qquad \partial_\sigma y_1=0,
\end{equation}
we can impose Dirichlet conditions for $y_2$ and Neumann for $y_1$ for any arbitrary $\sigma_D$.

Combining appropriately the above two cases, we can impose the following boundary conditions on the solutions:
\begin{itemize}
\item N-N for both $y_1$ and $y_2$ at $\sigma_i=\sigma_0$ and $\sigma_f=\sigma_0+\sqrt{\frac{1-\l}{1+\l}}\frac{\pi}{m}$, corresponding to a space-filling p2-brane.
\item D-D for $y_2$ and N-N $y_1$ at $\sigma_i$ and $\sigma_f$, which correspond to a pair of p1-branes.
\item D-N/N-D for $y_2$ and N-N $y_1$ at $\sigma=\sigma_D$ and $\sigma=\sigma_N$, corresponding to a p1-brane and a  p2-brane.
\end{itemize}
These solutions correspond to finite moving line segments.

\subsection{Boundary conditions - Analysis: 2nd class}
The second class of solutions reveals a much larger variety of results including static and translationally invariant configurations. It consists of fourteen distinct types of solutions. We keep the presentation as short as possible, presenting only basic facts for each of these types of solutions, but the overall presentation is lengthy.

\subsubsection{$SL(2,\mathbb{R})/U(1)_{V}$ Model}

 \subsubsection*{Static}
These solutions are of the form
\begin{equation}
y_1=\frac{1}{c}\sqrt{\frac{1-\lambda}{1+\lambda}}\cosh\left(m\,c\,\tau\right)\cos\psi(\sigma),\quad
y_2=\frac{1}{c}\sqrt{\frac{1+\lambda}{1-\lambda}}\cosh\left(m\,c\,\tau\right)\sin\psi(\sigma),
\end{equation}
where $\psi$ is given by \eqref{eq:psi_jacobi} and the corresponding elliptic modulus by \eqref{eq:elliptic_modulus_sl2}. The world-sheet of such solutions is cylindrical.

\begin{figure}[htb]
\centering
\includegraphics[width=0.4\textwidth]{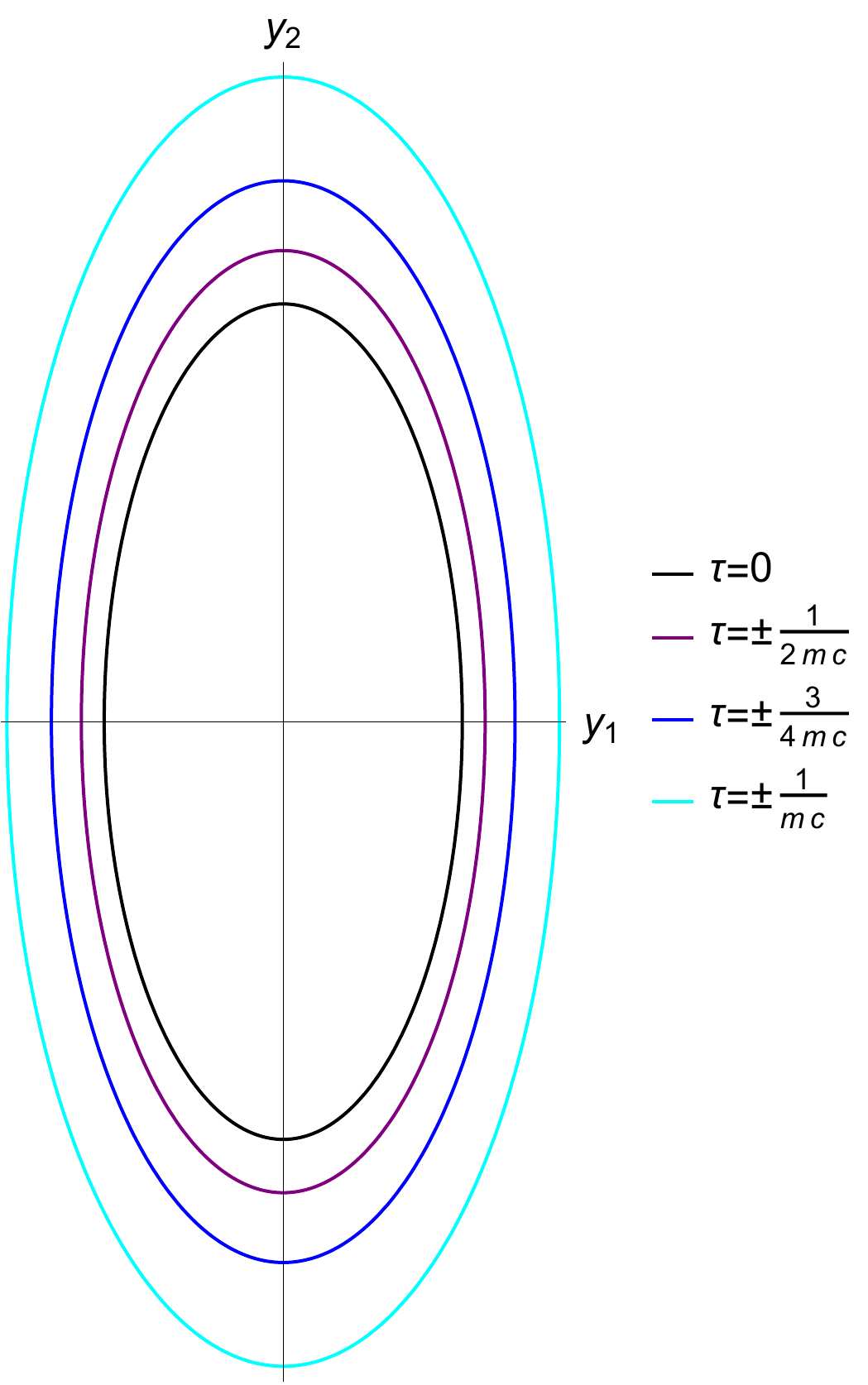}\includegraphics[width=0.4\textwidth]{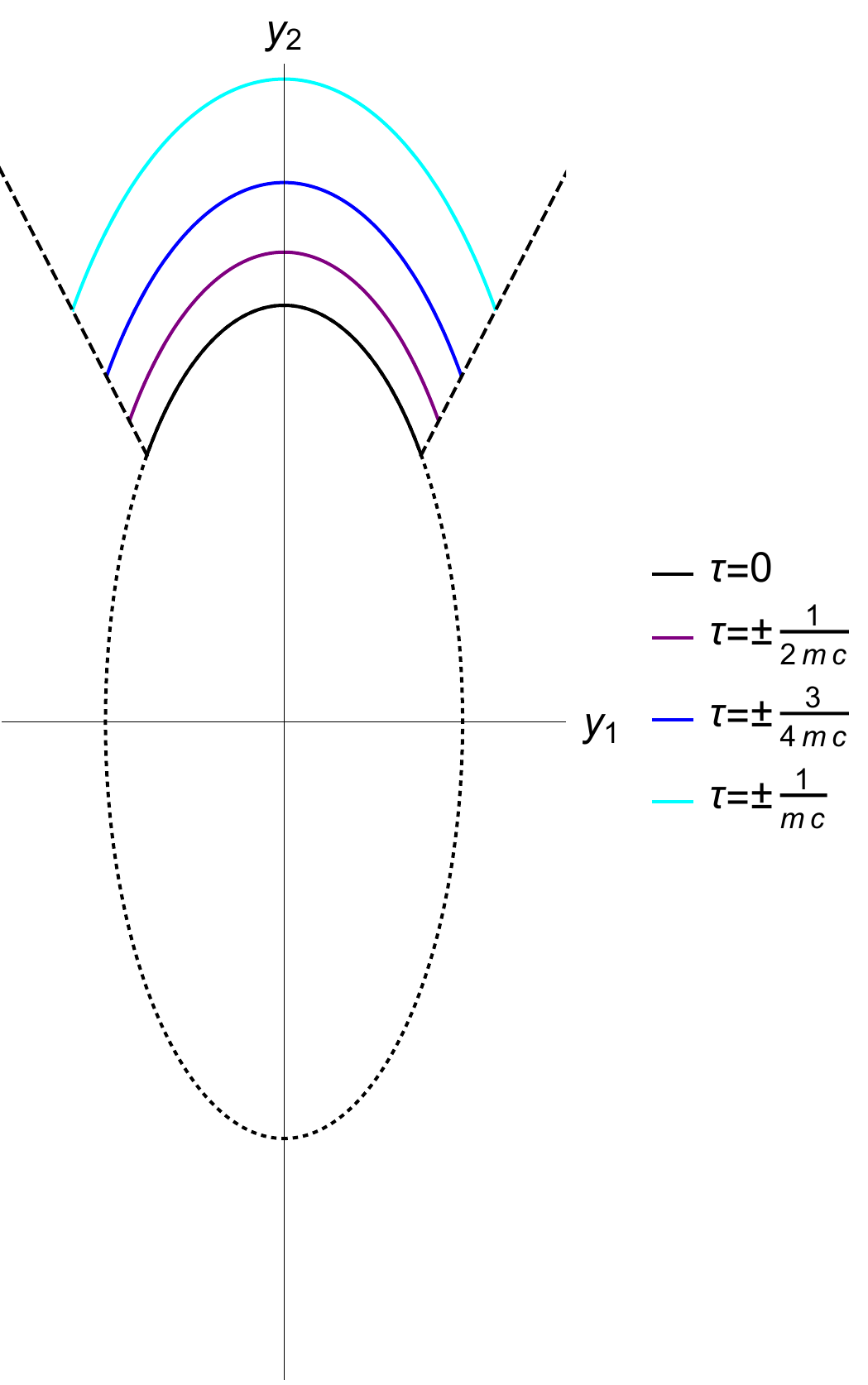}
\caption{Indicative static solutions of the $SL(2,\mathbb{R})/U(1)_V$ model. The left panel corresponds to a rotating solution, while the right one to an oscillating solution, which is a folded string. Notice that as $\lambda>0$ the angle $\psi$ of the oscillating solution ``oscillates'' around $\pi/2$. The equation of the black ellipses is \eqref{eq:boundary_ellipsis}.}
\label{fig:SL2V_static}
\end{figure}

\paragraph{Closed Strings}
For rotating solutions, the angle $\psi$ is monotonous and satisfies
\begin{equation}\label{eq:psi_quasiperiodicity}
\psi(\sigma+4\delta\sigma)=\psi(\sigma)+2\pi,
\end{equation}
just like the angle of a rotating pendulum, depicted in the right panel of Figure \eqref{fig:Jacobi_Amplitude}. Static rotating solutions correspond to closed strings. These solutions are ellipses, which, starting from  infinity, approach the origin of the $y_1$-$y_2$ plane up to $\tau=0$, and reflect back to infinity. Their motion, is bounded in the exterior of the ellipsis
\begin{equation}\label{eq:boundary_ellipsis}
y_1=\frac{1}{c}\sqrt{\frac{1-\lambda}{1+\lambda}}\cos\omega,\qquad y_2=\frac{1}{c}\sqrt{\frac{1+\lambda}{1-\lambda}}\sin\omega,\qquad \omega\in[0,2\pi)\,.
\end{equation}
The background is a deformed version of the well-known trumpet geometry. Intuitively, it has the same features but its cross-section is of elliptic shape. As the cross-section of the trumpet grows towards the origin $\rho=0$, the string approaching the origin stretches. At some point there is no more kinetic energy to be absorbed for the string to keep stretching, thus, it reflects back.

In the case of oscillating solutions, the angle $\psi$ ``oscillates'' between two limiting values and satisfies
\begin{equation}\label{eq:psi_periodicity}
\psi(\sigma+4\delta\sigma)=\psi(\sigma),
\end{equation}
where $\delta\sigma$ is given by \eqref{eq:ds_definition}, just like the angle of an oscillating pendulum depicted on the left panel of Figure \eqref{fig:Jacobi_Amplitude}. Intuitively, their motion is analogous to the rotating case, but the reason which prevents the string from collapsing to a points is not  topological, but kinematic. The endpoints of the string move at the speed of light. Indeed, it is easy to show that
\begin{equation}
G_{y_1 y_1}\left(\partial_{t}y_1\right)^2+G_{y_2 y_2}\left(\partial_{t}y_2\right)^2\vert_{\sigma=(2n+1)\delta\sigma}=1,\quad \forall t,
\end{equation}
where the time $t$ of the target space, defined as $t=m\tau$. Figure \ref{fig:SL2V_static} depicts indicative examples of closed static oscillating and rotating strings.

\paragraph{Open Strings}
As mentioned, in the case of rotating strings the angle $\psi$ is monotonous. Thus, in order to apply Dirichlet - Neumann (or Neumann - Dirichlet) boundary conditions, the only possibility is to set $\psi=n{\pi}/{2}$, where $n\in\mathbb{N}$. This is achieved for $\sigma=n\delta\sigma$. In this way we obtain strings, which are parts of the closed rotating strings ending on $y_1=0$ or $y_2=0$ axis. One can construct configurations, which extend along one, two or three quadrants and either end on different branes or on the same one.

In the case of oscillating strings, the angle $\psi$ ``oscillates'' either around $0$ or around $\pi/2$, depending on whether $\lambda>0$ or $\lambda<0$, see \eqref{eq:psi_jacobi}. The points of the string corresponding to $\sigma=n\delta\sigma$ either lie on an axis, or on the lines which are tangential to the motion of the folded string, like the dashed lines in the right panel of Figure \ref{fig:SL2V_static}.

\subsubsection*{Translationally Invariant}
These solutions are of the form
\begin{equation}
y_1=\frac{1}{c}\sqrt{\frac{1-\lambda}{1+\lambda}}\cosh\left(m\,c\,\sigma\right)\cos\psi(\tau),\quad
y_2=\frac{1}{c}\sqrt{\frac{1+\lambda}{1-\lambda}}\cosh\left(m\,c\,\sigma\right)\sin\psi(\tau),
\end{equation}
where $\psi$ is given by \eqref{eq:psi_jacobi} and the corresponding elliptic modulus by \eqref{eq:elliptic_modulus_sl2}. The world-sheet of such solutions is cylindrical.

\begin{figure}[htb]
\centering
\includegraphics[width=0.4\textwidth]{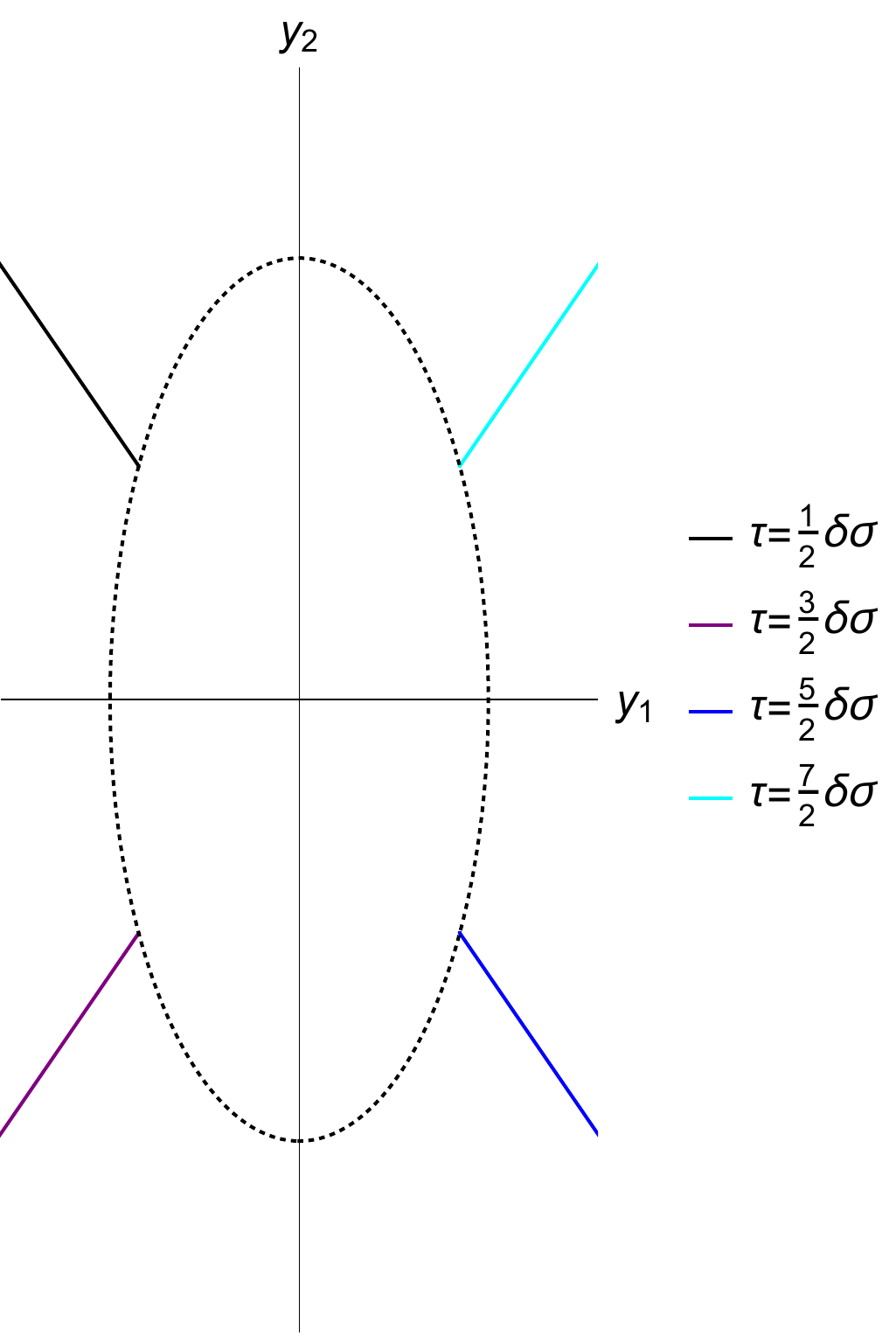}\includegraphics[width=0.4\textwidth]{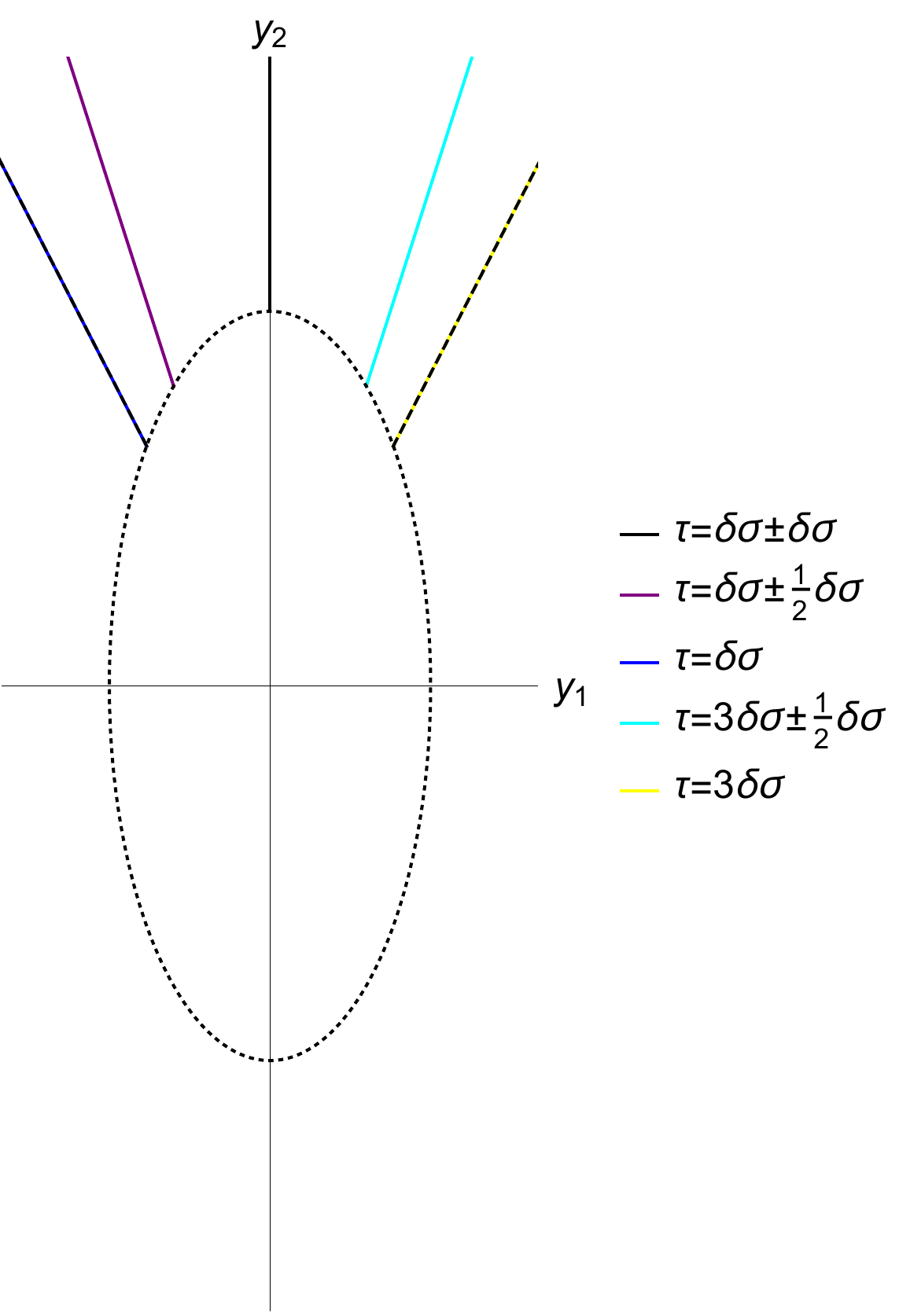}
\caption{Indicative translationally invariant solutions of the $SL(2,\mathbb{R})/U(1)_V$ model. The left panel corresponds to a rotating solution, while the right one to an oscillating solution. Notice that as $\lambda>0$ the angle $\psi$ of the oscillating solution ``oscillates'' around $\pi/2$. The equation of the dotted ellipses is  \eqref{eq:boundary_ellipsis}.}
\label{fig:SL2V_ti}
\end{figure}

These strings satisfy identically
\begin{equation}
G_{y_1 y_1}\left(\partial_{t}y_1\right)^2+G_{y_2 y_2}\left(\partial_{t}y_2\right)^2\vert_{\sigma=0}=1,\quad \forall t,
\end{equation}
where the time $t$ of the target space is defined as $t=m\tau$.

One can consider these configurations either as folded strings, for $\sigma\in(-\infty,\infty)$, in order to make the superficial terms vanish, or as open ones. As the $\sigma=0$ point moves at the speed of light, the string is prevented from collapsing. In the case of rotating strings, this point rotates on an elliptic trajectory, whereas in the case of oscillating strings, this point oscillates between two limiting points, see Figure \ref{fig:SL2V_ti}. In both cases the motion is periodic with period $T=4\delta\sigma$.

\subsubsection{$SL(2,\mathbb{R})/U(1)_{A}$ Model}
\subsubsection*{Static 1}
The static solutions of this class are of the form
\begin{equation}
y_1=\frac{1}{c}\sqrt{\frac{1-\lambda}{1+\lambda}}\sinh\left(m\,c\,\tau\right)\cos\psi(\sigma),\quad
y_2=\frac{1}{c}\sqrt{\frac{1+\lambda}{1-\lambda}}\sinh\left(m\,c\,\tau\right)\sin\psi(\sigma),
\end{equation}
where $\psi$ is given by \eqref{eq:psi_jacobi_ax1} and the corresponding elliptic modulus by \eqref{eq:elliptic_modulus_sl2}. The world-sheet of such solutions is cylindrical.

\begin{figure}[htb]
\centering
\includegraphics[width=0.4\textwidth]{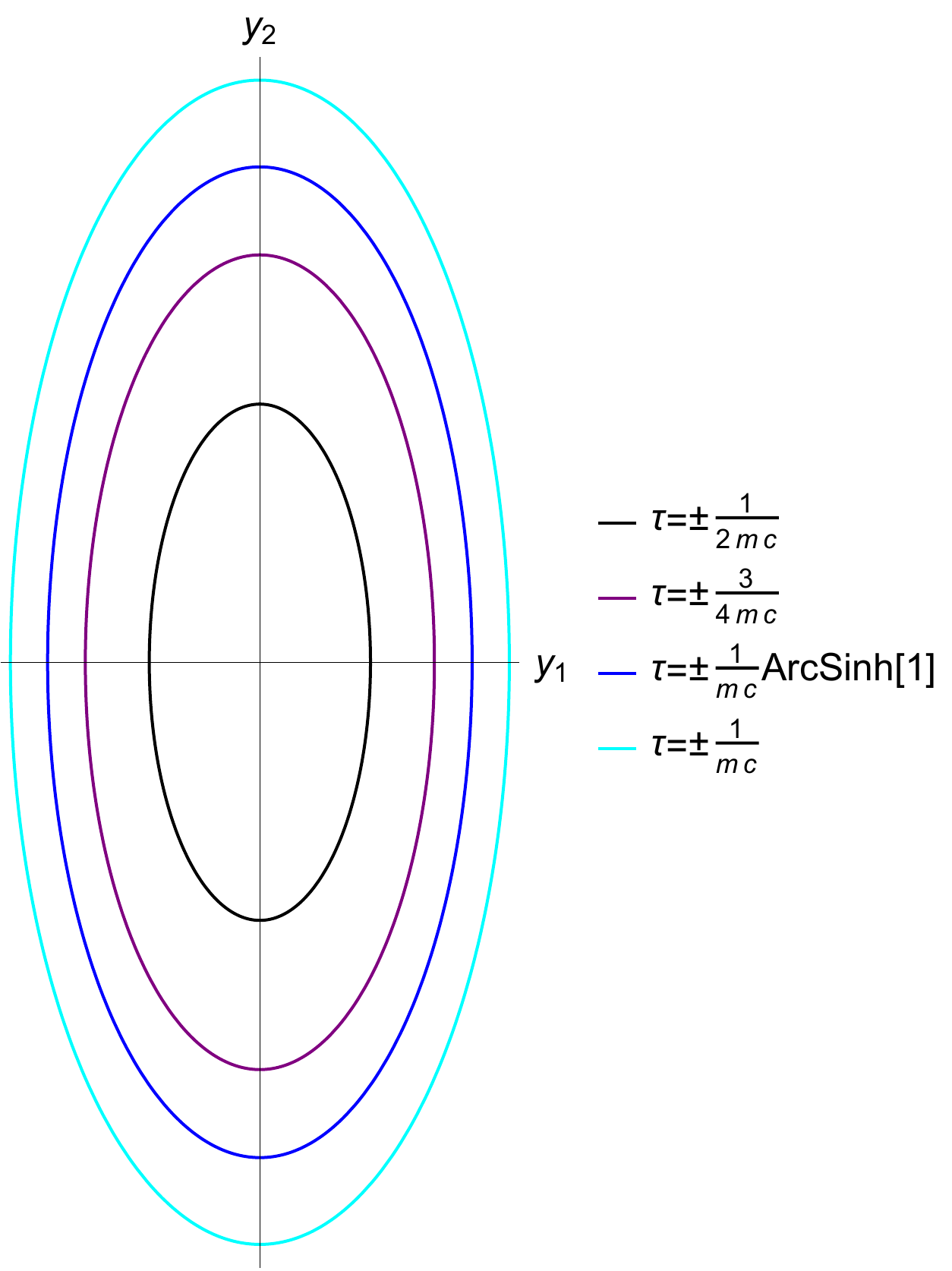}\includegraphics[width=0.4\textwidth]{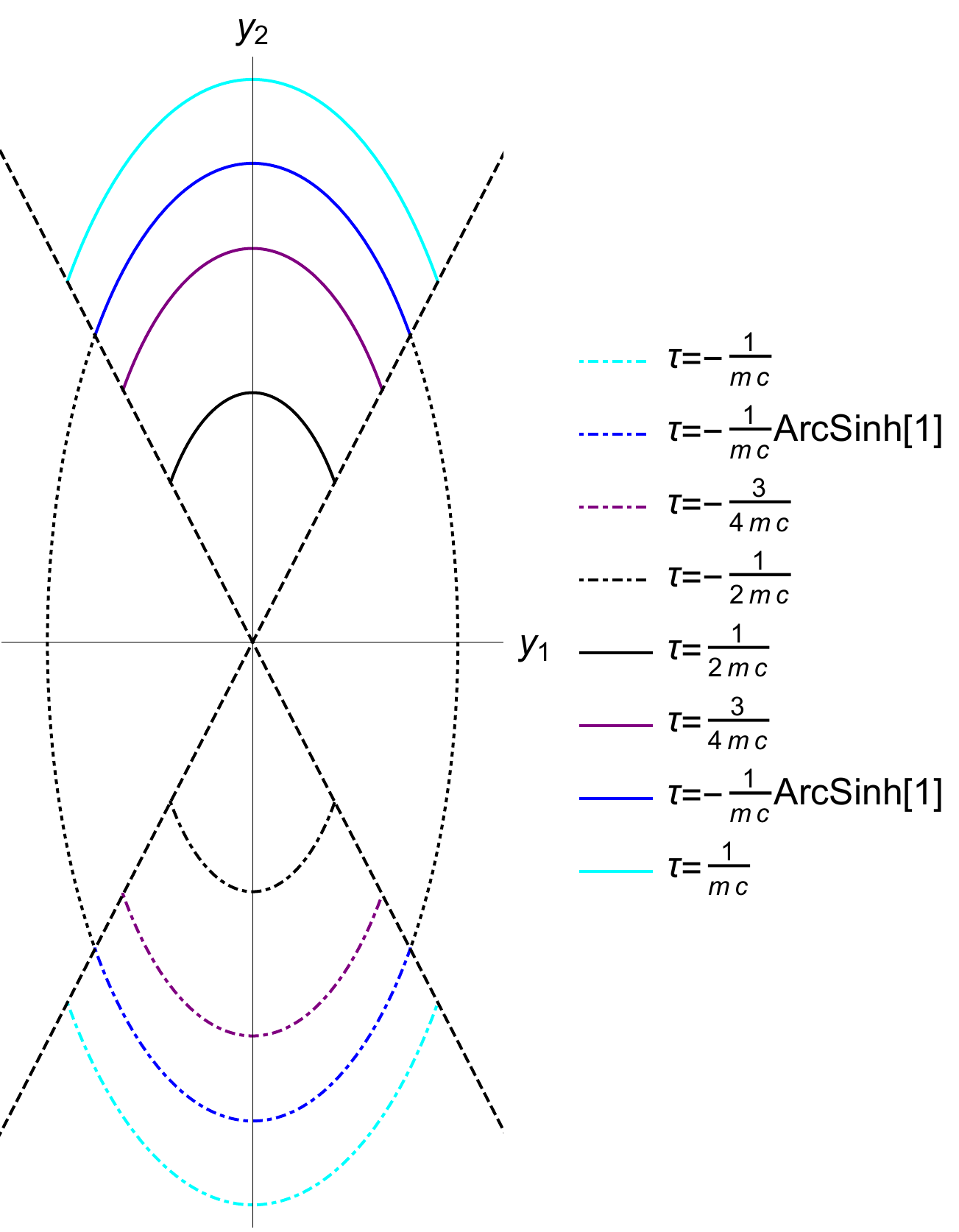}
\caption{Indicative static solutions of the $SL(2,\mathbb{R})/U(1)_A$ model. The left panel corresponds to a rotating solution, while the right one to an oscillating one, which is a folded string. Notice that as $\lambda>0$ the angle $\psi$ of the oscillating solution ``oscillates'' around $\pi/2$. The equation of the blue ellipses is\eqref{eq:boundary_ellipsis}.}
\label{fig:SL2A_static}
\end{figure}

\paragraph{Closed Strings}
The angle $\psi$ satisfies \eqref{eq:psi_quasiperiodicity} and \eqref{eq:psi_periodicity} in the case of rotating and oscillating solutions, respectively. Similarly to the case of the vectorially gauged model, the rotating solutions are ellipses. The background is a deformed version of Witten's cigar geometry \cite{Witten:1991yr}. Intuitively, it has the same characteristics but its cross-section is of elliptic shape. As the cross-section of the cigar shrinks towards the origin $\rho=0$, the string approaching the origin loosens. The incoming string becomes point-like at the tip of the cigar and then it is reflected back to infinity.

In the case of oscillating solutions, the motion of the strings is analogous to the rotating ones. Again the reason, which prevents the string from collapsing to a points is not the topological, but kinematic. The endpoints of the string move at the speed of light.
Figure \ref{fig:SL2A_static} depicts indicative examples of closed static oscillating and rotating strings.

\paragraph{Open Strings}
As in the case of the vectorially gauged model, we can enforce Dirichlet - Neumann (or Neumann - Dirichlet) boundary conditions to the rotating solutions, for $\sigma=n\delta\sigma$. In this way we obtain strings, which are parts of the closed rotating strings and end at the axis $y_1=0$ or $y_2=0$. One can construct configurations, which extend along one, two or three quadrants and either end on different branes or on the same one.

A similar picture emerges in the the case of oscillating strings. The angle $\psi$ ``oscillates'' either around $0$ or around $\pi/2$, depending on whether $\lambda>0$ or $\lambda<0$. The points of the string corresponding to $\sigma=n\delta\sigma$ either lie on an axis, or on the lines which are tangential to the motion of the folded string, the dashed lines in the right panel of Figure \ref{fig:SL2A_static}.

\subsubsection*{Translationally Invariant 1}
These solutions are of the form
\begin{equation}
y_1=\frac{1}{c}\sqrt{\frac{1-\lambda}{1+\lambda}}\sinh\left(m\,c\,\sigma\right)\cos\psi(\tau),\quad
y_2=\frac{1}{c}\sqrt{\frac{1+\lambda}{1-\lambda}}\sinh\left(m\,c\,\sigma\right)\sin\psi(\tau),
\end{equation}
where $\psi$ is given by \eqref{eq:psi_jacobi_ax1} and the corresponding elliptic modulus by \eqref{eq:elliptic_modulus_sl2}. The world-sheet of such solutions is cylindrical. 

\begin{figure}[htb]
\centering
\includegraphics[width=0.4\textwidth]{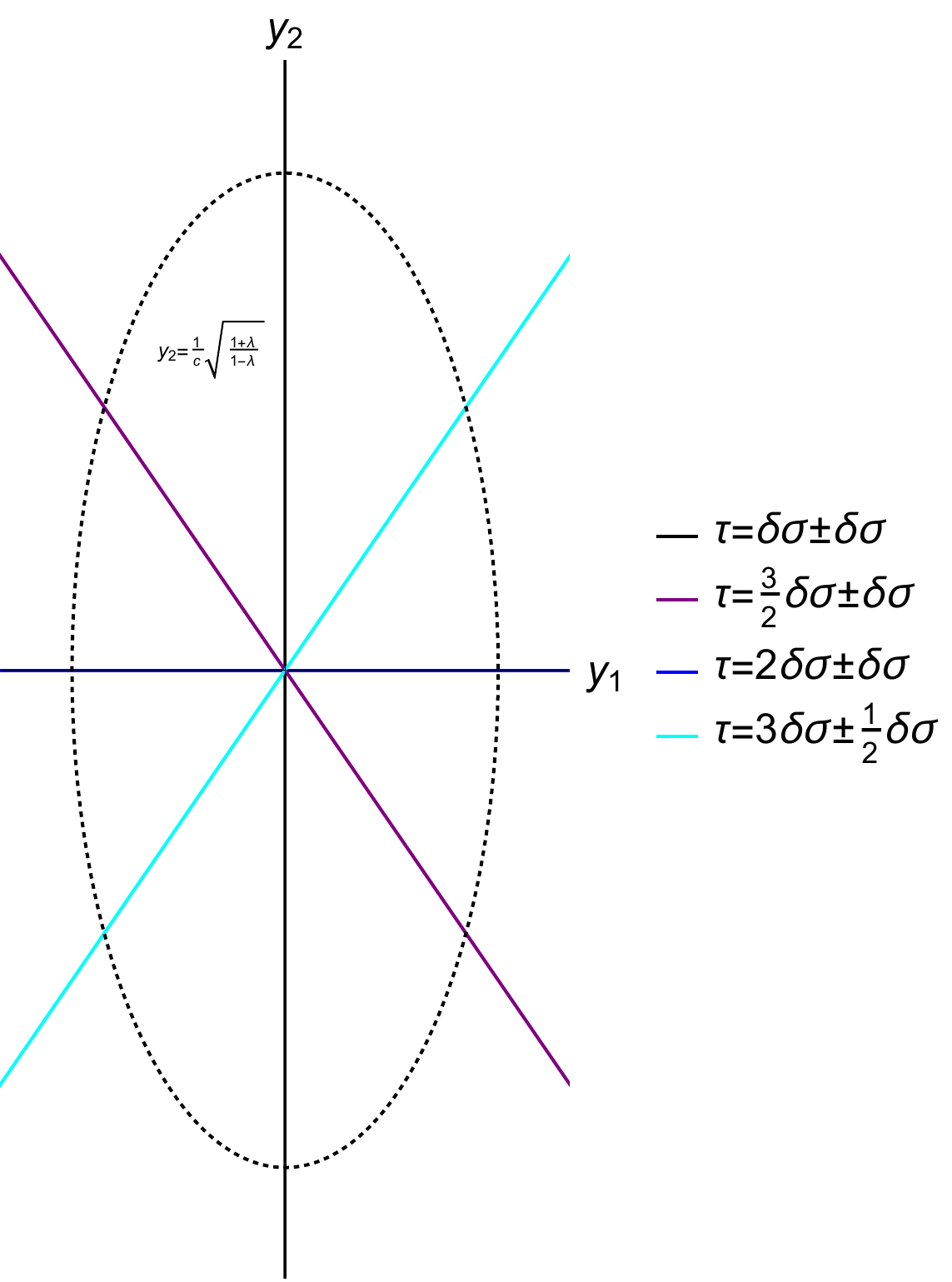}\includegraphics[width=0.4\textwidth]{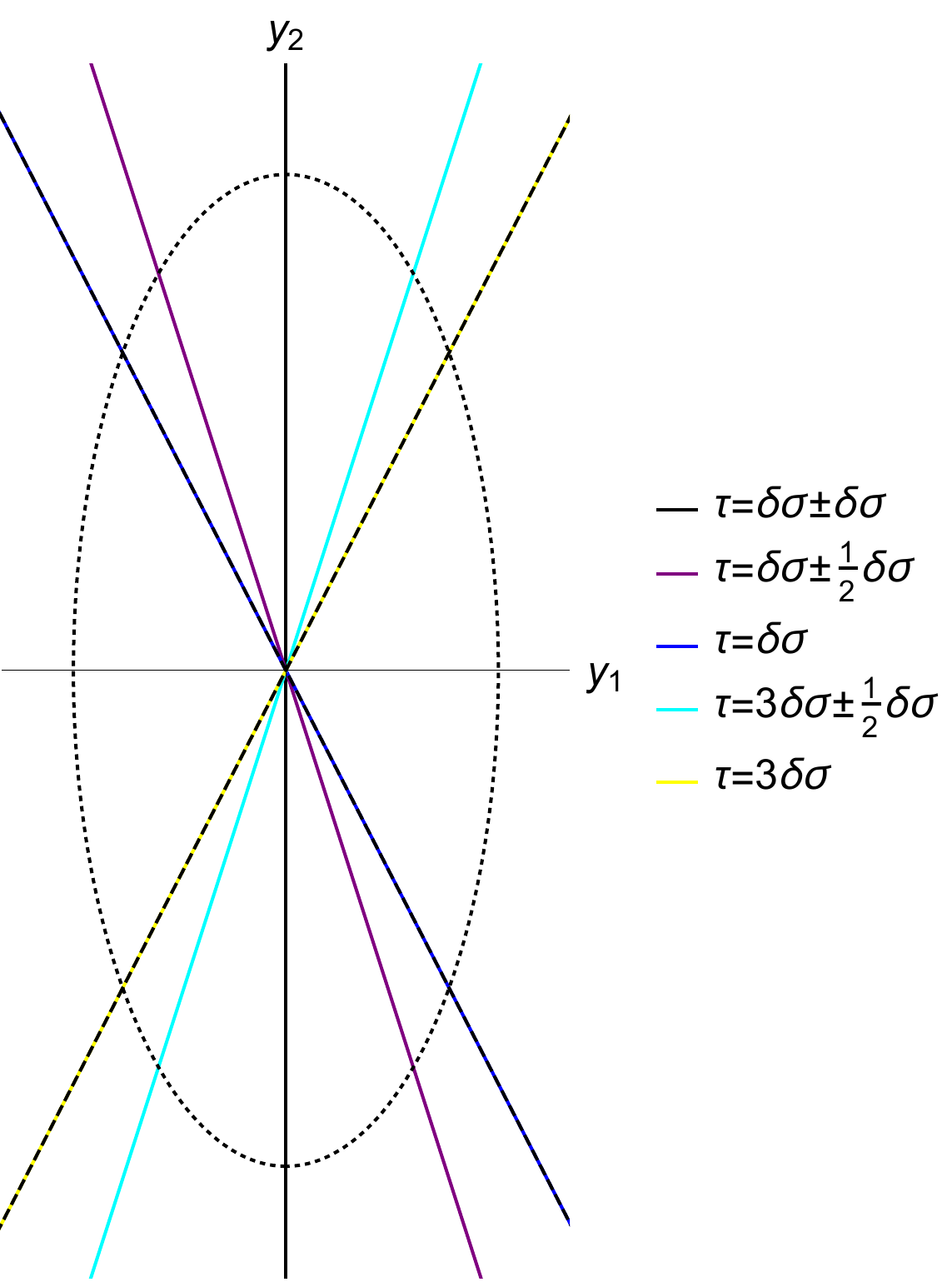}
\caption{Indicative translationally invariant solutions of the $SL(2,\mathbb{R})/U(1)_A$ model. The left panel corresponds to a rotating solution, while the right one to an oscillating solution. Notice, again, that as $\lambda>0$ the angle $\psi$ of the oscillating solution ``oscillates'' around $\pi/2$. The equation of dotted ellipses is \eqref{eq:boundary_ellipsis}.}
\label{fig:SL2A_ti}
\end{figure}

These strings satisfy identically
\begin{equation}
G_{y_1 y_1}\left(\partial_{t}y_1\right)^2+G_{y_2 y_2}\left(\partial_{t}y_2\right)^2\vert_{\sigma=0}=0,\quad \forall t,
\end{equation}
where the time $t$ of the target space is defined as $t=m\tau$. One can consider these configurations either as infinite strings, for $\sigma\in(-\infty,\infty)$, or semi-inifinite open ones, for $\sigma\in(0,\infty)$.  In the second case, the string is considered to end on a D0 brane, which is located at the tip of the cigar. This way the string is prevent from collapsing to a point.

In the case of rotating strings the string rotates freely, whereas in the case of oscillating strings, the string oscillates between two limiting points. The configurations are like the vectorially gauged ones in Figure \ref{fig:SL2V_ti}, but either with the strings ending on the origin of the plot, or extending all the way to infinity. They are depicted in Figure \ref{fig:SL2A_ti}. In both cases the motion is periodic with period $T=4\delta\sigma$.

\subsubsection*{Static 2}
These solutions are of the form
\begin{equation}
y_1=\frac{1}{c}\sqrt{\frac{1-\lambda}{1+\lambda}}\sin\left(m\,c\,\tau\right)\cos\psi(\sigma),\quad
y_2=\frac{1}{c}\sqrt{\frac{1+\lambda}{1-\lambda}}\sin\left(m\,c\,\tau\right)\sin\psi(\sigma),
\end{equation}
where $\psi$ is given by \eqref{eq:psi_jacobi_ax2} and the corresponding elliptic modulus by \eqref{eq:elliptic_modulus_tilde}. The angle $\psi$ of these solutions obeys equation \eqref{eq:psi_quasiperiodicity}, but in terms of $\delta\tilde{\sigma}$, defined in \eqref{eq:ds_tilde_definition}, instead of  $\delta\sigma$. There are only rotating solutions of this form. The world-sheet of such solutions is toroidal.

\begin{figure}[htb]
\centering
\includegraphics[width=0.40\textwidth]{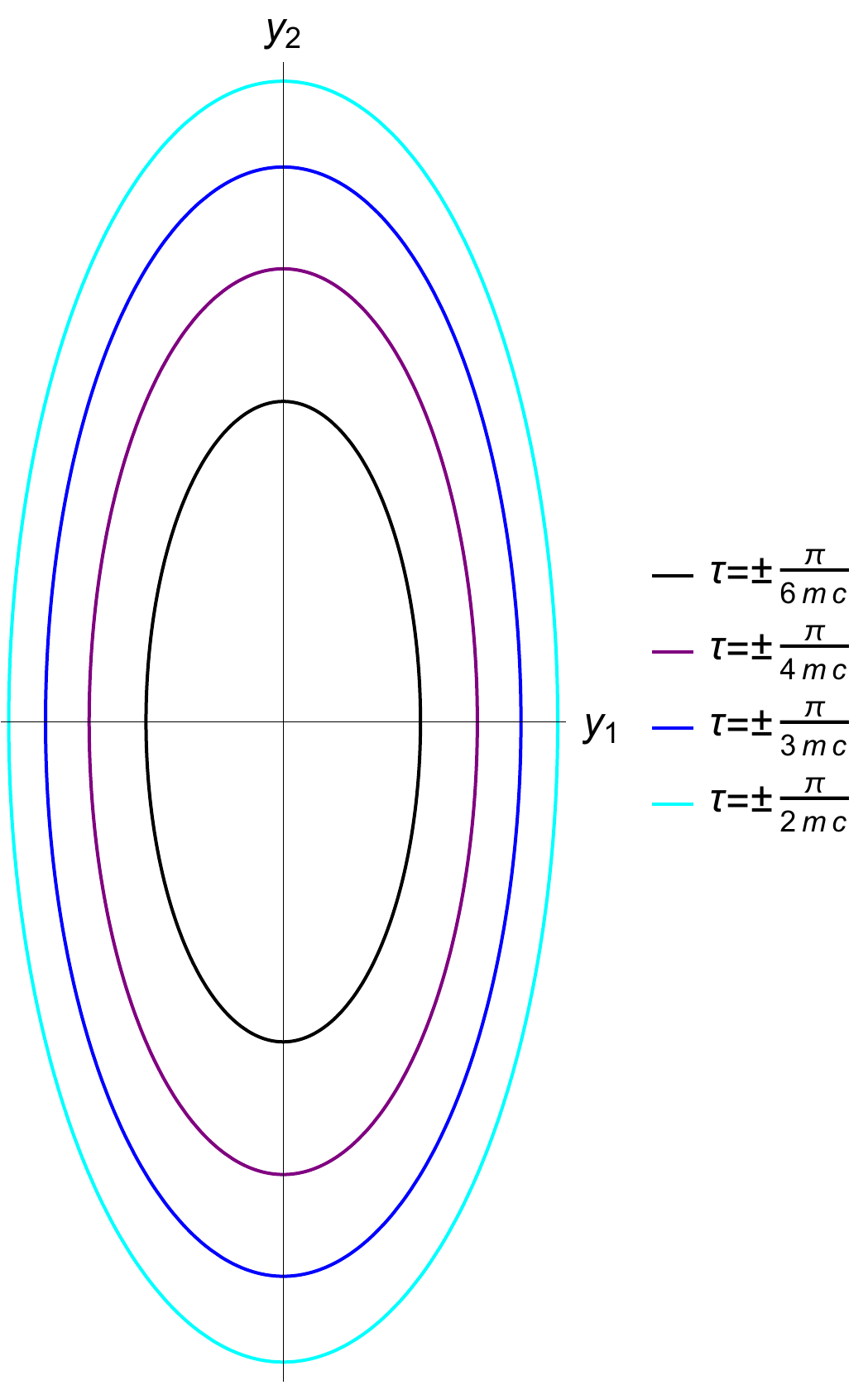}\includegraphics[width=0.40\textwidth]{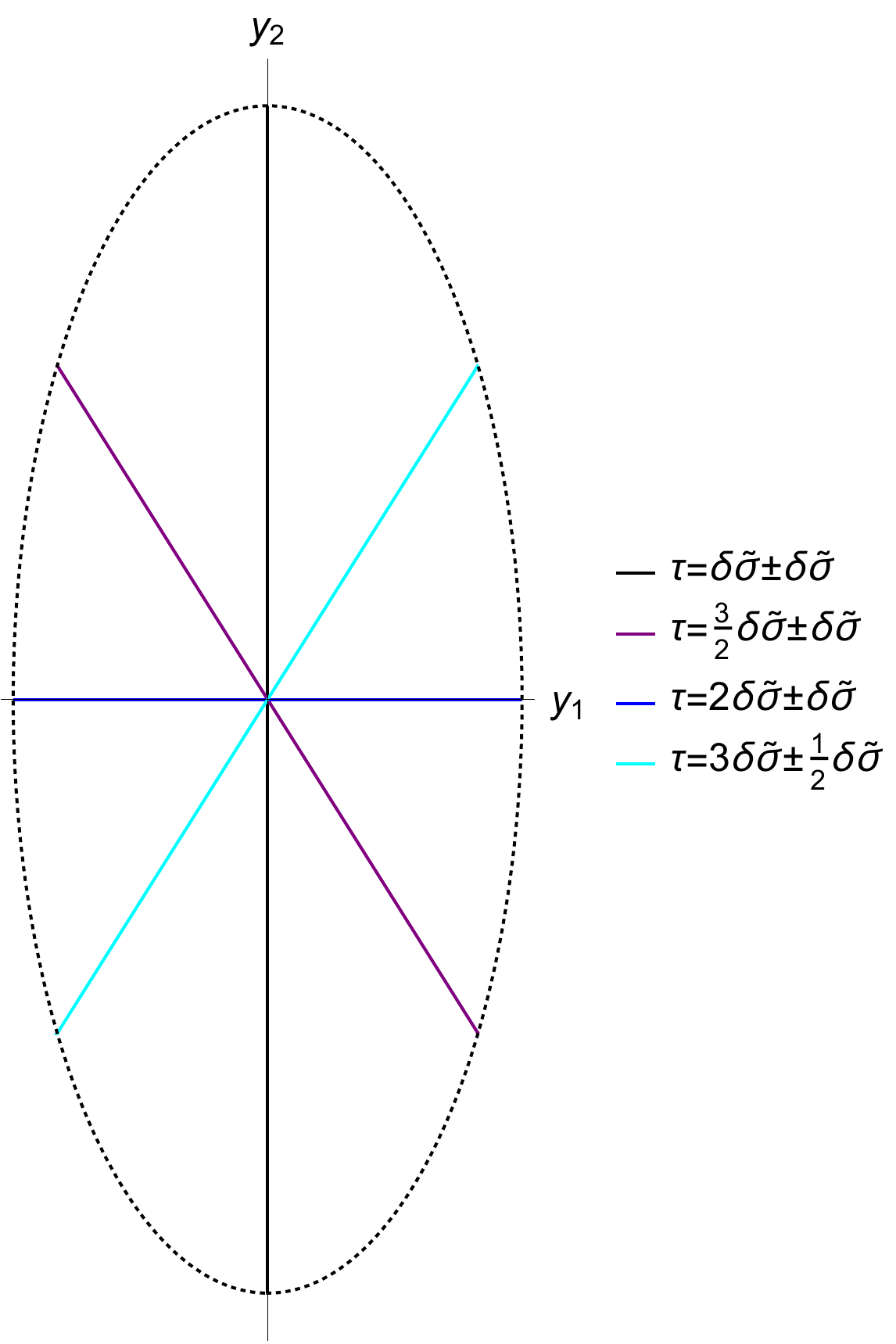}
\caption{Indicative rotating solutions of the $SL(2,\mathbb{R})/U(1)_A$ model. The left panel corresponds to a static solution, while the right one to a translationally invariant one, which is a folded string. The equation of the cyan ellipsis on the left panel and the dotted one on the right is \eqref{eq:boundary_ellipsis}.}
\label{fig:SL2A2}
\end{figure}

\paragraph{Closed Strings}
Similar to the static solutions presented so far, these solutions are ellipses. Again, the background is a deformed version of the Witten's cigar geometry \cite{Witten:1991yr}. Their motion is analogous to the static rotating solutions of the vectorially gauged model. Their motion is bounded by the same ellipsis, see equation \eqref{eq:boundary_ellipsis}, but these solutions move in the interior rather than the exterior. The string stretches until there is no more kinetic energy left and reflects back to the tip of the cigar. Such solutions are periodic both on the time-like and space-like world-sheet coordinates. The left panel of Figure \ref{fig:SL2A2} depicts indicative examples of such closed static rotating strings.

\paragraph{Open Strings}
As in the case of the vectorially gauged model, we can enforce Dirichlet - Neumann (or Neumann - Dirichlet) boundary conditions to the rotating solutions, for $\sigma=n\delta\tilde{\sigma}$. This way we obtain strings, which are parts of the closed rotating strings and end at axis $y_1=0$ or $y_2=0$. One can construct configurations, which extend along one, two or three quadrants and either end on different branes or on the same one.

\subsubsection*{Translationally Invariant 2}
The static solutions of this class are of the form
\begin{equation}
y_1=\frac{1}{c}\sqrt{\frac{1-\lambda}{1+\lambda}}\sin\left(m\,c\,\sigma\right)\cos\psi(\tau),\quad
y_2=\frac{1}{c}\sqrt{\frac{1+\lambda}{1-\lambda}}\sin\left(m\,c\,\sigma\right)\sin\psi(\tau),
\end{equation}
where $\psi$ is given by \eqref{eq:psi_jacobi_ax2} and the corresponding elliptic modulus by \eqref{eq:elliptic_modulus_tilde}. There are only rotating solutions of this form. The world-sheet of such solutions is toroidal.

These strings satisfy identically
\begin{align}
G_{y_1 y_1}\left(\partial_{t}y_1\right)^2+G_{y_2 y_2}\left(\partial_{t}y_2\right)^2\vert_{\sigma=0}=0,\quad \forall t,\\
G_{y_1 y_1}\left(\partial_{t}y_1\right)^2+G_{y_2 y_2}\left(\partial_{t}y_2\right)^2\vert_{\sigma=\pm\frac{1}{mc}\frac{\pi}{2}}=1,\quad \forall t,
\end{align}
where the time $t$ of the target space is defined as $t=m\tau$. These solutions correspond either to closed folded strings or to open ones, whose endpoint move at the speed of light. The motion of the endpoints prevents the string from shrinking to a point. One can also consider open strings which end on a D0-brane at the tip of the cigar. The right panel of Figure \ref{fig:SL2A2} depicts an indicative example of such closed translationally invariant rotating strings.

\subsubsection{$SU(2)/U(1)$ Model}
\subsubsection*{Static}
These solutions are of the form
\begin{equation}
y_1=\frac{1}{c}\sqrt{\frac{1-\lambda}{1+\lambda}}\sin\left(m\,c\,\tau\right)\cos\psi(\sigma),\quad
y_2=\frac{1}{c}\sqrt{\frac{1+\lambda}{1-\lambda}}\sin\left(m\,c\,\tau\right)\sin\psi(\sigma),
\end{equation}
where $\psi$ is given by \eqref{eq:solution_psi_am_su2} and the corresponding elliptic modulus by \eqref{eq:elliptic_modulus_bar}. The world-sheet of such solutions is toroidal.

\begin{figure}[htb]
\centering
\includegraphics[width=0.4\textwidth]{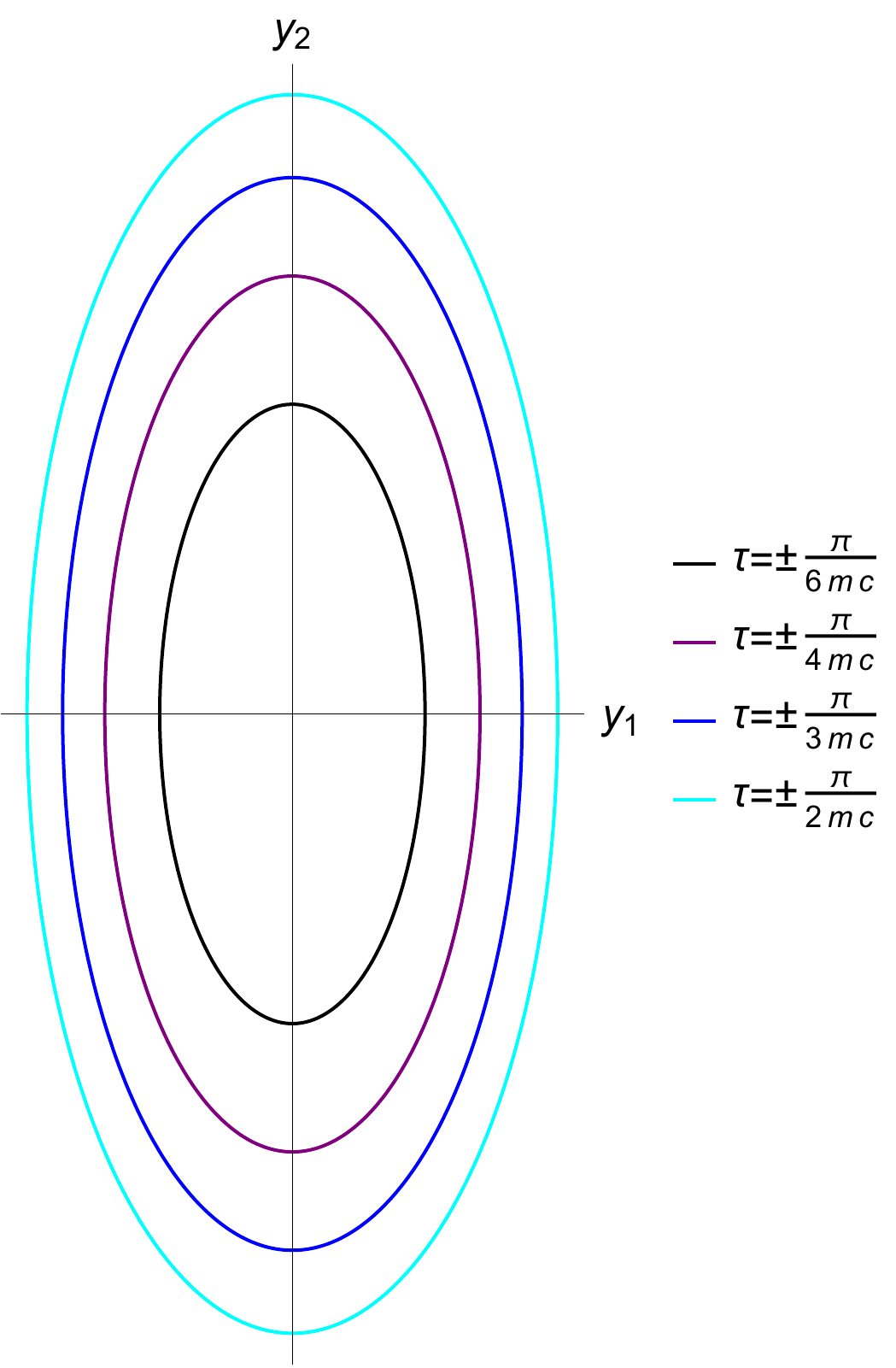}\includegraphics[width=0.4\textwidth]{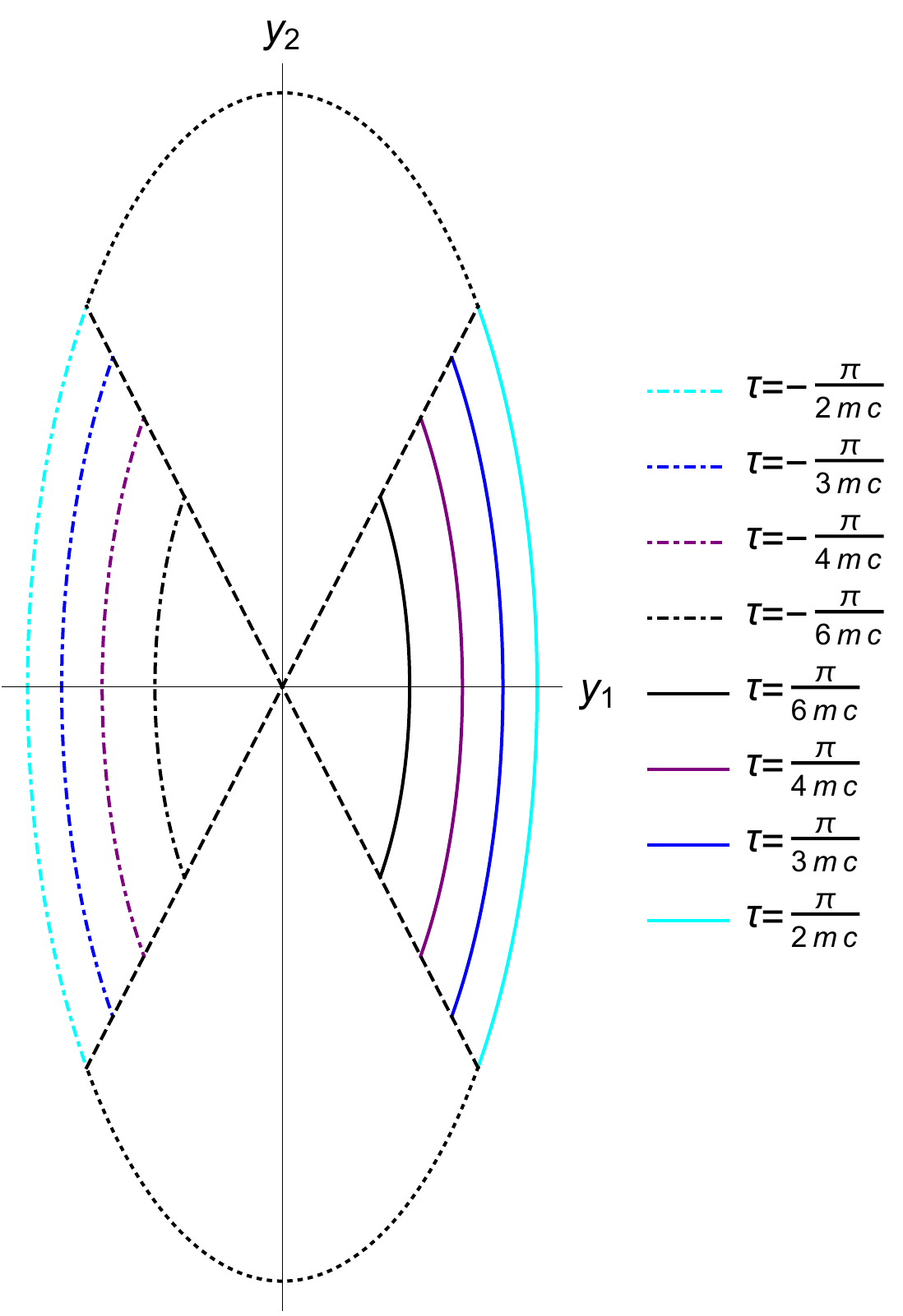}
\caption{Indicative static solutions of the $SU(2)/U(1)$ model. The left panel corresponds to a rotating solution, while the right one to an oscillating one, which is a folded string. Notice that as $\lambda>0$ the angle $\psi$ of the oscillating solution ``oscillates'' around $0$. The equation of the cyan ellipses is \eqref{eq:boundary_ellipsis}.}
\label{fig:SU2_static}
\end{figure}

\paragraph{Closed Strings}
The angle $\psi$ satisfies \eqref{eq:psi_quasiperiodicity} and \eqref{eq:boundary_ellipsis} in the case of rotating and oscillating solutions, respectively, but in terms of $\delta\bar{\sigma}$, defined in \eqref{eq:ds_bar_definition}, instead of $\delta\sigma$.

Static rotating solutions correspond to closed strings. They are ellipses, whose scale oscillates. Their motion,  is bounded by the ellipsis \eqref{eq:boundary_ellipsis}, which is expected as the manifold is compact. Considering the target space embedded in three dimensions, it has the shape of an spheroid. The string shrinks and stretches as it oscillates from a one pole to the other.

The oscillating solutions are folded strings. They are part of an ellipsis and their endpoints move at the speed of light. Their motion is analogous to the rotating ones, but in this case the reason, which prevents the string from collapsing to a points is not the topological, but kinematic. It is easy to show that
\begin{equation}
G_{y_1 y_1}\left(\partial_{t}y_1\right)^2+G_{y_2 y_2}\left(\partial_{t}y_2\right)^2\vert_{\sigma=(2n+1)\delta\sigma}=1,\quad \forall t,
\end{equation}
where the time $t$ of the target space, defined as $t=m\tau$.
Figure \ref{fig:SU2_static} depicts indicative examples of closed static oscillating and rotating strings.

\paragraph{Open Strings}
In the case of rotating strings the angle $\psi$ is monotonous. Thus, in order to enforce Dirichlet - Neumann (or Neumann - Dirichlet) boundary conditions, the only possibility is to set $\psi=n{\pi}/{2}$, where $n\in\mathbb{N}$. This is achieved for $\sigma=n\delta\bar{\sigma}$. This way we obtain strings, which are parts of the closed rotating strings and end at the axis $y_1=0$ or $y_2=0$. One can construct configurations, which extend along one, two or three quadrants and either end on different branes or on the same one.

In the case of oscillating strings, the angle $\psi$ ``oscillates'' either around $0$ or around $\pi/2$, depending on whether $\lambda>0$ or $\lambda<0$. The points of the string corresponding to $\sigma=n\delta\bar{\sigma}$ either lie on an axis, or on the lines which are tangential to the motion of the folded string, like the dashed lines in the right panel of Figure \ref{fig:SU2_static}.

\subsubsection*{Translationally Invariant}
These solutions are of the form
\begin{equation}
y_1=\frac{1}{c}\sqrt{\frac{1-\lambda}{1+\lambda}}\sin\left(m\,c\,\sigma\right)\cos\psi(\tau),\quad
y_2=\frac{1}{c}\sqrt{\frac{1+\lambda}{1-\lambda}}\sin\left(m\,c\,\sigma\right)\sin\psi(\tau),
\end{equation}
where $\psi$ is given by \eqref{eq:solution_psi_am_su2} and the corresponding elliptic modulus by \eqref{eq:elliptic_modulus_bar}. The world-sheet of such solutions is toroidal.
\begin{figure}[htb]
\centering
\includegraphics[width=0.4\textwidth]{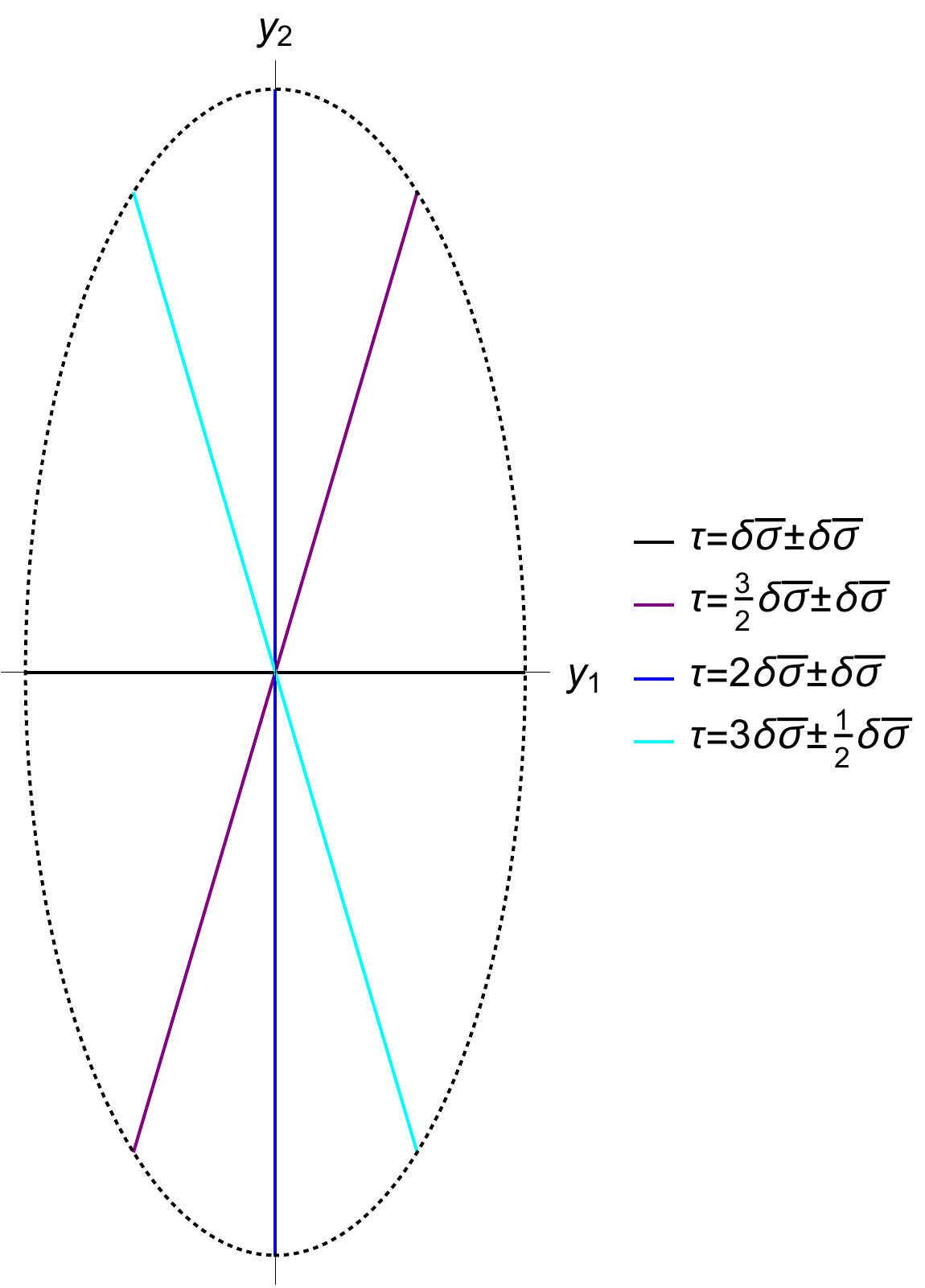}\includegraphics[width=0.4\textwidth]{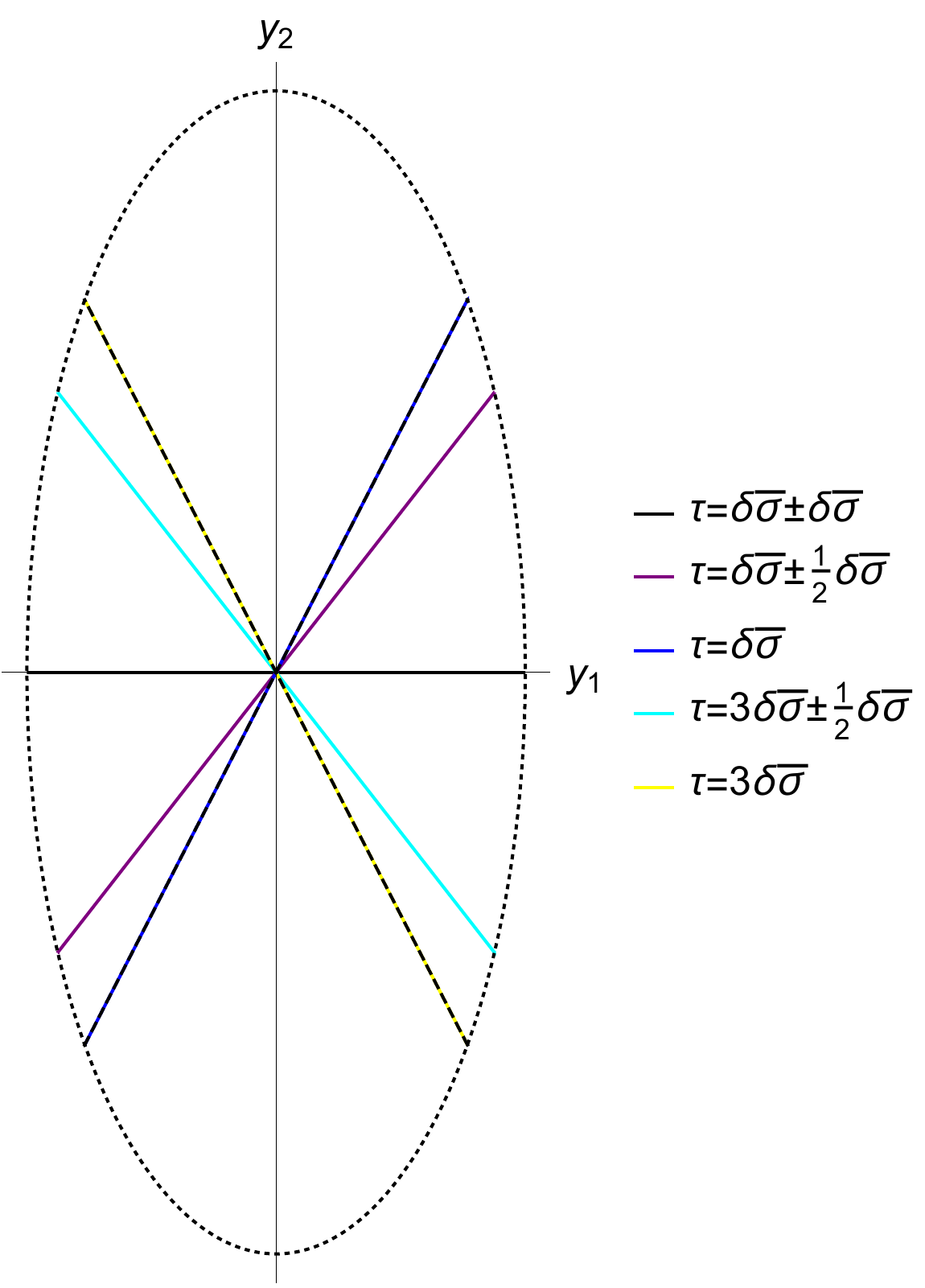}
\caption{Indicative translationally invariant solutions of the vectorially gauged $SU(2)/U(1)$ model. The left panel corresponds to a rotating solution, while the right one to an oscillating solution. Notice that as $\lambda>0$ the angle $\psi$ of the oscillating solution ``oscillates'' around $0$.}
\label{fig:SU2_ti}
\end{figure}
These strings satisfy identically
\begin{equation}
G_{y_1 y_1}\left(\partial_{t}y_1\right)^2+G_{y_2 y_2}\left(\partial_{t}y_2\right)^2\vert_{\sigma=\pm\frac{1}{mc}\frac{\pi}{2}}=1,\quad \forall t,
\end{equation}
where the time $t$ of the target space is defined as $t=m\tau$.

One can consider these configurations either as folded strings, or as open ones. As the endpoints moves at the speed of light, the string is prevent from collapsing. In the case of rotating strings, this point rotates on an elliptic trajectory, whereas in the case of oscillating strings, these points oscillate between two limiting ones, see \ref{fig:SU2_ti}. In both cases the motion is periodic with period $T=4\delta\bar{\sigma}$.

\subsection{The effect of the non-perturbative symmetries}
Regarding the first class of solutions, let us return for a moment in equations \eqref{y1} and \eqref{y2} and impose invariance under the duality symmetry \eqref{duality}. In order to do so, we have to postulate that $m^2$ is a function of $\l$ transforming as
\be
\label{m}
m^2\left(1/\lambda\right)=-m^2(\l).
\end{equation}
On the contrary, $c$ is not affected in implying that $c(1/\l)=c(\l)$.

As far as the duality \eqref{-lsymmetry} is concerned, one can easily see from \eqref{y1} and \eqref{y2}, that invariance under this duality implies that
\be
\label{-lcsymmetry}
 m^2(-\l)=m^2(\l)\qquad c(-\l)= -c(\l)
\end{equation}
We conclude that $c$, is even under \eqref{duality} and odd under \eqref{-lsymmetry}symmetry. The inverse holds for $m^2$ \footnote{ As an indicative example, one may define  
\begin{equation}
m^2=\frac{1-\lambda^2}{1+\lambda^2}m^2_0,\qquad c=\frac{\lambda}{1+\lambda^2}c_0,
\end{equation}
where $m_0$ and $c_0$ are invariant under both transformations.}.  One can easily verify that the sets of solutions in Tables \ref{tab:solutions_sl2_vector}, \ref{tab:solutions_sl2_axial} and \ref{tab:solutions_su2}, are closed under symmetries \eqref{duality} and \eqref{-lcsymmetry}.

Let us turn to the second class of solutions. In the case of the $SL(2,\mathbb{R})/U(1)$ model, the duality symmetry \eqref{duality} acts on the coordinates $\rho$ and $\phi$ as
\begin{equation}
\lambda\to 1/\lambda,\qquad \rho\to-\rho,\qquad \phi\to-\phi.
\end{equation}
Taking into account equations \eqref{chipsi1_ap}, \eqref{eq:def_psi}  and \eqref{eq:def_dilaton_r_phi}, we infer that $\chi$, $\psi$ and the dilaton $\tilde\Phi$ transform as
\be
\label{chipsiduality}
\chi\to \chi+i\p/2,\qquad\psi\to \psi, \qquad e^{2\tilde\Phi}\to\,\,-e^{2\tilde\Phi}.
\end{equation}
As a result, the action \eqref{eq:action_sl2_chi_psi_vector} is invariant under  \eqref{duality}.

In order to retain the duality symmetry at the level of solutions, equations \eqref{quadchipsi}, \eqref{eq:chi_eom_sl2} and \eqref{eq:psi_eom_sl2} imply that $m^2$ and $c^2$ are functions of $\l$ satisfying
\be
m^2\left(1/\l\right)=-m^2(\l)\qquad c^2\left(1/\l\right)=-c^2(\l).
\end{equation}
Equation \eqref{eq:elliptic_modulus_sl2} implies that $\kappa^2$ is invariant, thus \eqref{eq:psi_jacobi} implies that $\psi$ is indeed invariant. Finally, according to equation \eqref{eq:exp_chi}, $\chi$ transforms appropriately.

Last but not least, the symmetry \eqref{-lsymmetry}, in terms of coordinates $(\theta,\phi)$ is replaced by
\be
\label{thetaphi}
\rho\to \rho,\qquad\phi\to\phi+\p/2
\end{equation}
and one can show that under \eqref{thetaphi}
\be
\chi\to\chi,\qquad \psi\to \psi+\p/2,
\end{equation}
while 
\be
m^2(-\l)=m^2(\l),\qquad c^2(-\l)=c^2(\l).
\end{equation}
Similar conclusions hold for the case of axial gauging too, as well as for the $SU(2)/U(1)$ model.

\section{Discussion}
\label{sec:Discussion}
In this work we have derived as a first example in the literature, two distinct classes of solutions for the $SL(2,\mathbb{R})/U(1)$ and $SU(2)/U(1)$ $\l$-deformed manifolds. We achieved this be exploiting a special property of the 2-dimensional target spaces, namely the fact that the solutions of the Energy-Momentum tensor conservation equations  also solve the equations of motion.

The solutions of the first class are expressed in terms of trigonometric - hyperbolic functions. Overall we obtain eleven distinct solutions of this kind, which are gathered in Tables \ref{tab:solutions_sl2_vector}, \ref{tab:solutions_sl2_axial} and \ref{tab:solutions_su2}. All these configurations are of the form $y_1=y_1(\tau)$ and $y_2=y_2(\sigma)$, but also the $\sigma\leftrightarrow\tau$ transformed solutions are also valid. The solutions of the first class of the $SL(2,\mathbb{R})/U(1)$  $\l$-deformed models are infinite, semi-infinite or finite moving line segments, whereas the ones of the $SU(2)/U(1)$  $\l$-deformed model are finite line segments, which oscillate.

The solutions of the second class are expressed in terms of Jacobi's elliptic functions. Overall we obtain fourteen distinct solutions of this kind, which are gathered in Table \ref{tab:solutions_2nd}. In addition, Table \ref{tab:solutions_2nd_categories} present the classification of these solutions. These configurations are either of the form $y_1=f(\tau)\cos\psi(\sigma)$
and $y_2=f(\tau)\sin\psi(\sigma)$, which are referred to as static, or of the form $y_1=f(\sigma)\cos\psi(\tau)$ and $y_2=f(\sigma)\sin\psi(\tau)$,  which are referred to as translationally invariant. It is important to point out, that in this case the $\sigma\leftrightarrow\tau$ transformation relates solutions, which are not related by any target space symmetry. The static solutions are of elliptical shape, whose scale is time-dependent, whereas the translationally invariant solutions are semi-infinite rotating line segments, whose endpoint lies on an ellipses. The second class of solutions have two special limits. The first one is obtained when the elliptic modulus vanishes and the elliptic functions degenerate to trigonometric functions. This is the case when $\lambda=0$. For this class of solutions, the $\lambda-$deformation turns the trigonometric functions to elliptic ones. The second one is obtained for some specific value of $c$, which depends on the model. In this limit the elliptic functions degenerate to hyperbolic ones and the corresponding configurations are presented in Table \ref{tab:solutions_2nd_kink}.

Both configurations are not necessarily invariant under the non-perturbative duality symmetry, but this can be imposed as an extra demand. Solution profiles are time dependent and do not saturate a BPS bound. It remains obscured how these configurations are incorporated in the spectrum of the theory.

Several new directions emerge through this analysis. First of all, it would be interesting to study the Pohlmeyer reduced theory of the models \cite{Hollowood:2014rla}. Such a study may provide information relevant to the spectrum of the theory. For instance when the elliptic modulus of the second class of solutions equals to unity, are configurations are analogous to the Giant Magnon \cite{Hofman:2006xt} and one expects that the Pohlmeyer avatar is a solitonic object. 

On the same line, the derived solutions may be used in order to obtain new ones. This can be achieved via the application of the dressing method \cite{Zakharov:1973pp,Harnad:1983we}. The dressing method is a technique, which takes advantage of a known solution, usually refereed to as the seed solution, in order to obtain new ones. The advantage of this method is that in order to derive the solutions one has to solve a system of linear coupled first order PDEs, rather than the equations of motion, which constitute a system of coupled, non-linear, second order PDEs. This method has already been applied in the context of $\lambda-$deformations in \cite{Appadu:2017xku}. In this work the seed solution is analogous to the BMN particle \cite{Berenstein:2002jq}, a solution of the undeformed model, which also solves the deformed one. The solutions we derived are much more complicated and the application of the dressing on such seed solutions may reveal interesting phenomena, such as the formation of spikes and memory effect regarding the propagation of the inserted kink on the non-trivial background \cite{Katsinis:2019oox}. 

It is worth noticing that the dressing method is also related with the stability analysis of the seed solutions \cite{Katsinis:2019oox}. Obviously, it also has the advantage that besides determining the fate of small perturbations, one also obtains the full non-linear solution as well. Of course, even a linear stability analysis is of interest. Its conclusions are expected to match the ones of the dressing method \cite{Katsinis:2019sdo}.

It would, also be interesting to derive kink configurations i.e time-independent lumps of finite energy, or time independent solutions in general. Unfortunately, for such configurations the approach based on the Energy-Momentum conservation is inapplicable. This is also the case for models, whose target space is of higher dimension. Regarding models, 
generalizing the ones of this work, i.e. ones having the groups considered here as subgroups, one can obtain solutions via the dressing method. To do so, one needs to embed the solutions of this work in the higher-dimensional target space and apply the dressing method using this seed solution.

Besides further investigating the solutions themselves or the corresponding models, there are various studies, which are related to them. To begin with, it would be interesting to find classical solutions of other theories having a 2-dimensional target space using the approach based on the Energy-Momentum conservation. For instance, one may obtain solutions of various known integrable deformations, such as Yang Baxter \cite{Klimcik:2008eq} / $\eta$ \cite{Delduc:2013fga,Delduc:2013qra}, bi Yang Baxter \cite{Klimcik:2014bta} and asymptotic $\lambda$-deformations \cite{Itsios:2021wso}. Regarding the Yang Baxter and $\eta$ deformations, it is known that they are related to $\lambda-$deformations via Poisson-Lie T-duality and appropriate analytic continuations \cite{Sfetsos:2015nya}. Thus, one could test whether the solutions obtained here, also solve the $\eta$-deformed models after necessary manipulations. As far as asymptotic $\lambda$-deformations are concerned, it would be interesting to derive solutions for the theory corresponding to the asymptotic limit of the $SL(2,\mathbb{R})/U(1)$ model. In this case we expect a larger variety of results, as a consequence of symmetry enhancement. Such solutions may be relevant to the original FZZ-duality \cite{FZZ,Hikida:2008pe} or to the study of the effect of the $\lambda-$deformation on the duality. 

Moreover, as we described a lot of open string configurations, another interesting direction, is to investigate the fitting of D-branes in this setup and derive classical solutions for such objects. One can also study whether the boundary conditions preserve the integrability of the theory or not \cite{Driezen:2018glg}.  

Finally, the derived solutions enable various field theory calculations. In particular, one could study the effective theory related to a non-trivial classical solution and perform semi-classical quantization.

\section*{Acknowledgements}
The work of D.K. was supported by FAPESP grants 2019/21281-4 and 2021/01819-0. The authors want to thank Konstantinos Sfetsos, Konstantinos Siampos and George Georgiou for fruitful discussions and comments at several steps of the project. 

\appendix

\section{The Parametrization of the 2nd class of solutions}
\label{sec:parametrization}
In this appendix we  give some technical details on the second parametrization that we introduced in order to describe the models under study. 
\subsection*{$SL(2,\mathbb{R})/U(1)$ Vector Gauging}
Let us discuss the case of the vectorially gauged $SL(2,\mathbb{R})/U(1)$ model. Applying the change of variables \eqref{apcoor_sl2_vector}, the action \eqref{lWitten_sl2_vector} is written as
\be
\begin{split}\label{vectorgauged}
&S=\frac{k}{\p}\int\,d^2\s\,\Big\{\frac{1-\l}{1+\l}\left(\del_+\rho\del_-\rho +\coth^2\rho\del_+\phi\del_-\phi\right)\\
&\phantom{000000}+\frac{4\l}{1-\l^2}\left(\cos\phi\del_+\rho-\sin\phi\coth \rho\del_+\phi\right)\left(\cos\phi\del_-\rho-\sin\phi\coth \rho\del_-\phi\right)\Big\}.
\end{split}
\end{equation}
With some work one can show that defining the new coordinates $\chi$ and $\psi$ as
\begin{align}
\chi(\rho,\phi)&=\log\left[\cosh \rho\sqrt{\frac{1-\l}{1+\l}+\frac{4\l}{1-\l^2}\cos^2\phi}\right], \label{chipsi1_ap}\\
\psi(\rho,\phi)&=\arctan\left[\frac{1-\l}{1+\l}\tan\phi\right]. \label{eq:def_psi}
\end{align}
the action can be written in a conformally flat form, namely \eqref{eq:action_sl2_chi_psi_vector}, where initially the dilaton is defined as
\be
e^{2\tilde\Phi}=\coth^2 \rho\left(\frac{1-\l}{1+\l}+\frac{4\l}{1-\l^2}\cos^2\phi\right).\label{eq:def_dilaton_r_phi}
\end{equation}
Note that the square root in the definition of the $\chi$ variable, equation \eqref{chipsi1_ap}, should not worry us, because it is well defined for $-1<\lambda<1$, since
\be\label{eq:cos_inequality}
\frac{1+\vert\l\vert}{1-\vert\l\vert}\geq\,\,\frac{1-\l}{1+\l}+\frac{4\l}{1-\l^2}\cos^2\phi\,\,\geq\frac{1-\vert\l\vert}{1+\vert\l\vert},
\end{equation}
Of course, the dilaton field has to be expressed in terms of the new variables $(\chi,\psi)$. To invert \eqref{chipsi1_ap} and \eqref{eq:def_psi} we take advantage of the equation
\be\label{eq:phi_psi_relation}
\frac{1-\l}{1+\l}+\frac{4\l}{1-\l^2}\cos^2\phi=\left(\frac{1+\l}{1-\l}-\frac{4\l}{1-\l^2}\cos^2\psi\right)^{-1},
\end{equation}
which is a direct consequence of \eqref{eq:def_psi}. Doing so, we obtain following expressions
\begin{align}
\rho(\chi,\psi)&=\text{arccosh}\left[e^{\chi}\sqrt{\frac{1+\l}{1-\l}-\frac{4\l}{1-\l^2}\cos^2\psi}\right],\label{eq:r_vec}\\
\phi(\psi)&=\arctan\left[\frac{1+\l}{1-\l}\tan\psi\right]. \label{psi}
\end{align}
It order to specify uniquely $\psi$ in terms of $\phi$ we choose
\begin{equation}\label{eq:phi_to_psi}
\cos\psi=\frac{\cos\phi}{\sqrt{\left(\frac{1-\lambda}{1+\lambda}\right)^2+\frac{4\lambda}{\left(1+\lambda\right)^2}\cos^2\phi}}, \qquad\sin\psi=\frac{\frac{1-\lambda}{1+\lambda}\sin\phi}{\sqrt{\left(\frac{1-\lambda}{1+\lambda}\right)^2+\frac{4\lambda}{\left(1+\lambda\right)^2}\cos^2\phi}}.
\end{equation}
Taking into account \eqref{eq:phi_psi_relation}, the inverse transformation is
\begin{equation}\label{eq:psi_to_phi}
\cos\phi=\frac{\cos\psi}{\sqrt{\left(\frac{1+\lambda}{1-\lambda}\right)^2-\frac{4\lambda}{\left(1-\lambda\right)^2}\cos^2\psi}}, \qquad\sin\phi=\frac{\frac{1+\lambda}{1-\lambda}\sin\psi}{\sqrt{\left(\frac{1+\lambda}{1-\lambda}\right)^2-\frac{4\lambda}{\left(1-\lambda\right)^2}\cos^2\psi}}.
\end{equation}
These definitions are common for all three models.

In view of \eqref{eq:cos_inequality}, the quantity under the square root is positive provided that $\psi$ is a real function. Nevertheless, the validity of \eqref{eq:r_vec} requires
\be
\label{geq}
e^{\chi}\sqrt{\frac{1+\l}{1-\l}-\frac{4\l}{1-\l^2}\cos^2\psi}\,\,\,\geq1.
\end{equation}
This is a constraint that has to be imposed on the solutions. Implementing \eqref{eq:r_vec} and \eqref{psi} on the dilaton profile \eqref{eq:def_dilaton_r_phi}, we can finally express it in terms of $\chi$ and $\psi$ as
\be
\label{3.13}
e^{-2\tilde\Phi}=\frac{1+\l}{1-\l}-\frac{4\l}{1-\l^2}\cos^2\psi-e^{-2\chi}.
\end{equation}
Provided the inequality \eqref{geq} is satisfied, the dilaton $\tilde\Phi$ is real valued as required.

\subsection*{$SL(2,\mathbb{R})/U(1)$ Axial Gauging}
The vectorially and axially gauged actions of $SL(2,\mathbb{R})/U(1)$ are essentially related by the interchange $\sinh\rho\leftrightarrow \cosh\rho$. This means that the action of the axially gauged model is provided by \eqref{vectorgauged} upon the substitution $\coth\rho\to\tanh\rho$. Obviously the replacement discussed above, does not alter the definition of $\psi$, i.e. equation \eqref{eq:def_psi}, but the definition of $\chi$ should be adjusted appropriately. Thus, $\chi$ is given by
\begin{equation}\label{eq:def_chi_sl2_ax}
\chi(\rho,\phi)=\log\left[\sinh \rho\sqrt{\frac{1-\l}{1+\l}+\frac{4\l}{1-\l^2}\cos^2\phi}\right].
\end{equation}
Contrary to the case of vector gauging, the inverse transformation, namely
\begin{equation}\label{eq:r_axial}
\rho(\chi,\psi)=\text{arcsinh}\left[e^{\chi}\sqrt{\frac{1+\l}{1-\l}-\frac{4\l}{1-\l^2}\cos^2\psi}\right]
\end{equation}
is valid automatically and does not impose any constraint on the parameters. The first new entry here is the expression relating the fields $\chi$ and $\psi$ with the new dilaton field. Following the step of the previous case, the conformal factor is given in terms of $\chi$ and $\psi$ by
\be
\label{dilation_chi_psi_ax}
e^{-2\tilde\Phi}=\frac{1+\l}{1-\l}-\frac{4\l}{1-\l^2}\cos^2\psi+e^{-2\chi}\,\,\,.
\end{equation}

\subsection*{$SU(2)/U(1)$}
Using the analytic continuation \eqref{eq:analytic_continuation} relating the $SU(2)/U(1)$ to the vectorially gauged $SL(2,\mathbb{R})/U(1)$ in \eqref{vectorgauged}, the action of the former reads
\be
\begin{split}
&S=\frac{k}{\p}\int\,d^2\s\,\Big\{\frac{1-\l}{1+\l}\left(\del_+\theta\del_-\theta +\cot^2\theta\del_+\phi\del_-\phi\right)\\
&\phantom{000000}+\frac{4\l}{1-\l^2}\left(\cos\phi\del_+\theta+\sin\phi\cot \theta\del_+\phi\right)\left(\cos\phi\del_-\theta+\sin\phi\cot \theta\del_-\phi\right)\Big\}.
\end{split}
\end{equation}
The action can be written in the conformally flat form form \eqref{eq:action_sl2_chi_psi_vector} using
\begin{equation}
\chi(\theta,\phi)=\frac{1}{2}\log\left[\cos^2 \theta\left(\frac{1-\l}{1+\l}+\frac{4\l}{1-\l^2}\cos^2\phi\right)\right],\label{eq:def_chi_su2}
\end{equation}
and \eqref{eq:def_psi}, while the new dilaton field is given by
\be
e^{-2\tilde\Phi}= e^{-2\chi} -\left(\frac{1+\l}{1-\l}-\frac{4\l}{1-\l^2}\cos^2\psi\right).
\end{equation}
Implementing \eqref{eq:phi_psi_relation}, one can show that $\theta$ is given by
\be
\theta=\text{arccos}\left(e^{\chi}\sqrt{\frac{1+\l}{1-\l}-\frac{4\l}{1-\l^2}\cos^2\psi}\right).\label{eq:theta}
\end{equation}
Actually, this equation determines the absolute value of $\cos\theta$. One should define $\theta$, so that it is continuous and smooth. The validity of this equation requires
\begin{equation}
1\geq e^{\chi}\sqrt{\frac{1+\l}{1-\l}-\frac{4\l}{1-\l^2}\cos^2\psi}\geq -1.
\end{equation}
This inequality guaranties that the dilaton is real valued.

\section{The Jacobi Elliptic Functions}
\label{sec:Jacobi_review}
In this section we gather some properties of the Jacobi elliptic functions, which are relevant for this work. The fundamental object of Jacobi elliptic functions is the Jacobi amplitude $\mathrm{am}(x\vert m)$. Essentially, is generalizes the linear function $f(x)=x$. Using the Jacobi amplitude, one defines the elliptic sine and cosine as
\begin{equation}\label{eq:Jacobi_sn_cn}
\mathrm{sn}(x|m)=\sin\left(\mathrm{am}(x|m)\right),\qquad \mathrm{cn}(x|m)=\cos\left(\mathrm{am}(x|m)\right).
\end{equation}
The Jacobi amplitude satisfies the differential equation\footnote{The reader should be aware that it is quite common to find this differential equations in the form \begin{equation*}
\left(\frac{d}{dx} \mathrm{am}(x\vert m)\right)^2=1-m^2 \mathrm{sn}^2(x\vert m),
\end{equation*}
In this work we follow the conventions of Wolfram Mathematica.}
\begin{equation}\label{eq:Jacobi_ode}
\left(\frac{d}{dx} \mathrm{am}(x\vert m)\right)^2=1-m \,\mathrm{sn}^2(x\vert m),
\end{equation}
where $m$ is called the elliptic modulus. The second class of solutions is obtained using this equation. Trivially, it follows that
\begin{equation}\label{eq:vacuum_limit}
\mathrm{am}(x\vert 0)=x.
\end{equation}
Additionally, the Jacobi amplitude obeys
\begin{equation}
\mathrm{am}(0\vert m)=0.
\end{equation}
In this work, the elliptic modulus is always positive, so we set $m=\kappa^2$, where $\kappa\in\mathbb{R}$. Depending on whether $0<\kappa^2<1$ or $1<\kappa^2$ the Jacobi amplitude is either periodic or quasi-periodic function. In the special case $\kappa^2=1$ it follows that
\begin{equation}\label{eq:kink_limit}
\mathrm{am}(x\vert 1)=2\arctan\left(e^x\right)-\frac{\pi}{2}.
\end{equation}
The function is neither periodic or quasi-periodic and interpolates monotonically from $-\pi/2$ to $\pi/2$. This is the famous kink of the sine-Gordon equation. In this case the elliptic functions degenerate to hyperbolic ones, namely
\begin{equation}
\mathrm{sn}(x|1)=\tanh(x),\qquad \mathrm{cn}(x|1)=\mathrm{sech}(x).
\end{equation}

In order to proceed, let us consider a simple pendulum. This physical system will reveal all features of the Jacobi amplitude. The energy conservation of the pendulum reads
\begin{equation}\label{eq:Pendulum_Energy}
\frac{1}{2}\left(\frac{d\phi}{dt}\right)^2+\omega^2\left(1-\cos\phi\right)=E.
\end{equation}
The energy is normalized so that the (stable) equilibrium point $\phi=0$ corresponds to $E=0$. Trivially, this equation assumes the form 
\begin{equation}
\frac{1}{2E}\left(\frac{d\phi}{dt}\right)^2=1-\frac{2\omega^2}{E}\sin^2\frac{\phi}{2},
\end{equation}
which is solved by
\begin{equation}
\phi(t)=2\textrm{am}\left(\sqrt{\frac{E}{2}}(t-t_0)\big\vert \frac{2\omega^2}{E}\right)
\end{equation}

If $E<2\omega^2$, the pendulum oscillates between the angles $-\phi_0$ and $\phi_0$, where the value of the angle $\phi_0$ is $\phi_0=\arcsin\left(\sqrt{\frac{E}{2\omega^2}}\right)$. This behaviour is general, if $\kappa^2>1$, the Jacobi amplitude is bounded. In particular, it follows that
\begin{equation}
-\arcsin(1/\kappa)\leq \textrm{am}\left(x\vert\kappa^2\right)\leq \arcsin(1/\kappa),\qquad \kappa\geq 1.
\end{equation}
If $E>2\omega^2$, the motion of the pendulum is no longer oscillatory. The pendulum rotates, but its motion is modulated by the gravitational force. On average the angle grows linearly with time. Of course, as the energy grows, the modulation becomes less significant. Finally, if $E=2\omega^2$ the motion of the pendulum is aperiodic. Starting from the unstable equilibrium point $\phi=-\pi$, after an infinite amount of time, the pendulum reaches the unstable equilibrium point $\phi=\pi$. In the main text, solutions corresponding to $\kappa^2>1$ are referred to as oscillating, whereas solutions corresponding $0<\kappa^2<1$ as rotating.

\begin{figure}
\centering
\includegraphics[width=0.31\textwidth]{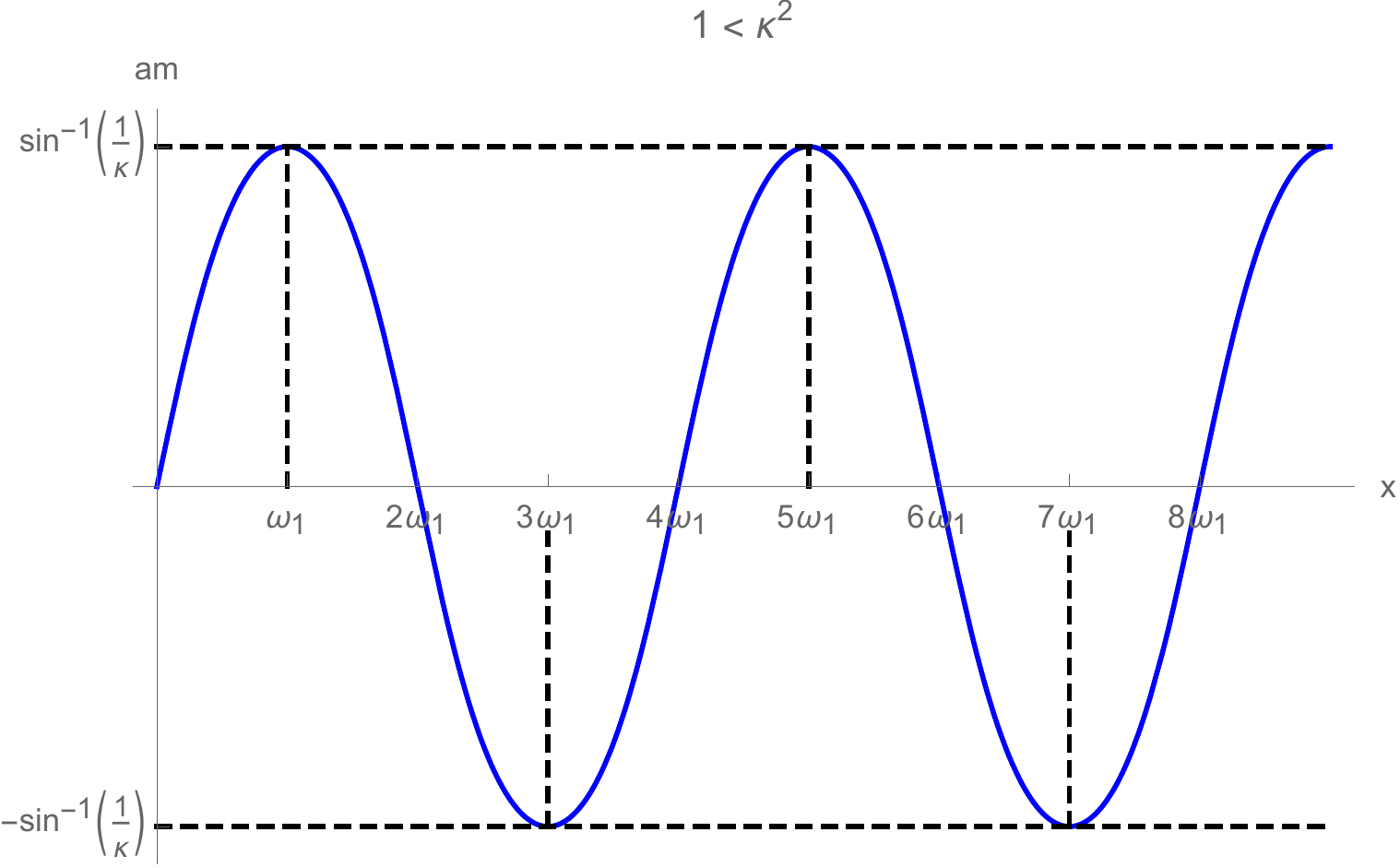}\includegraphics[width=0.31\textwidth]{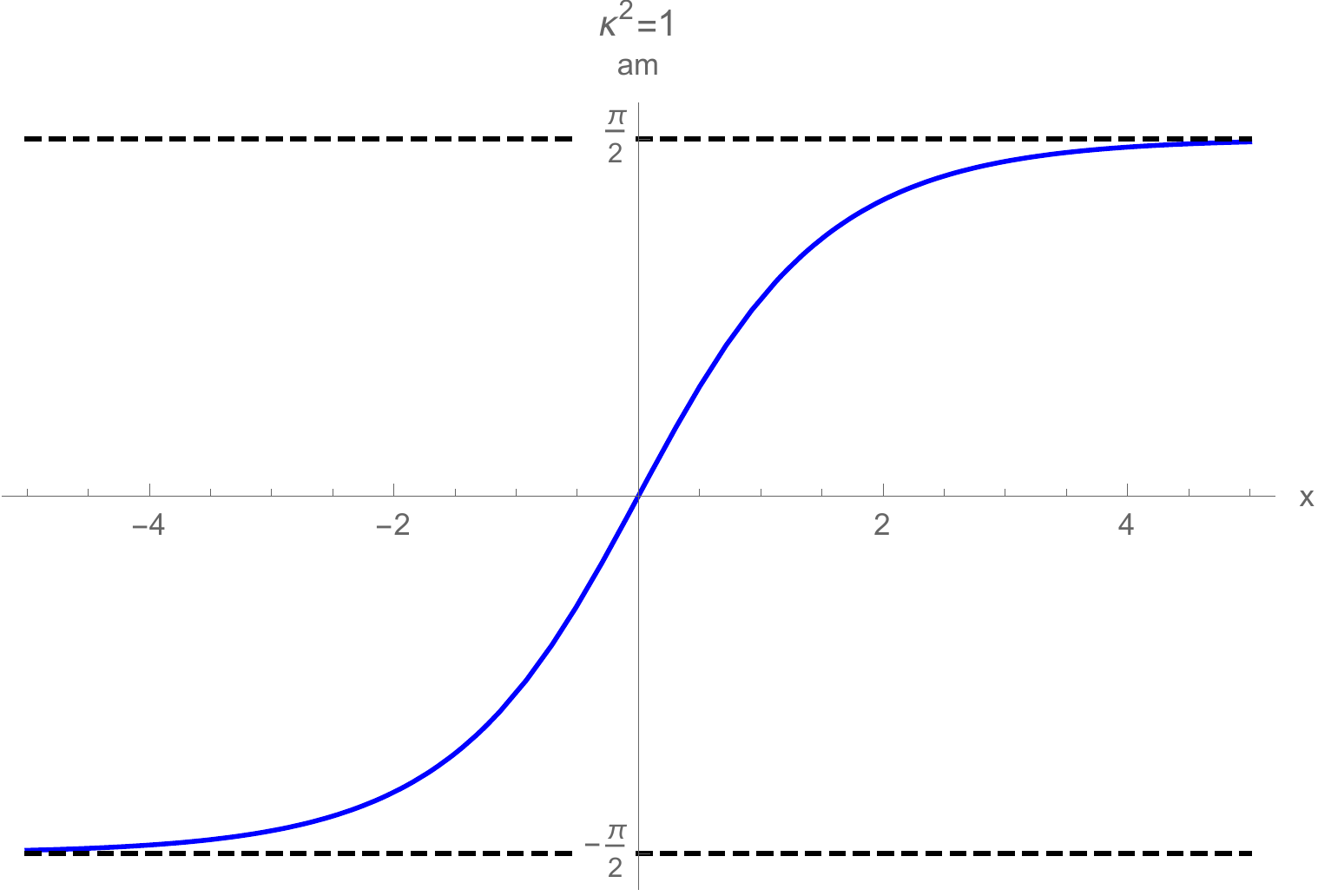}\includegraphics[width=0.31\textwidth]{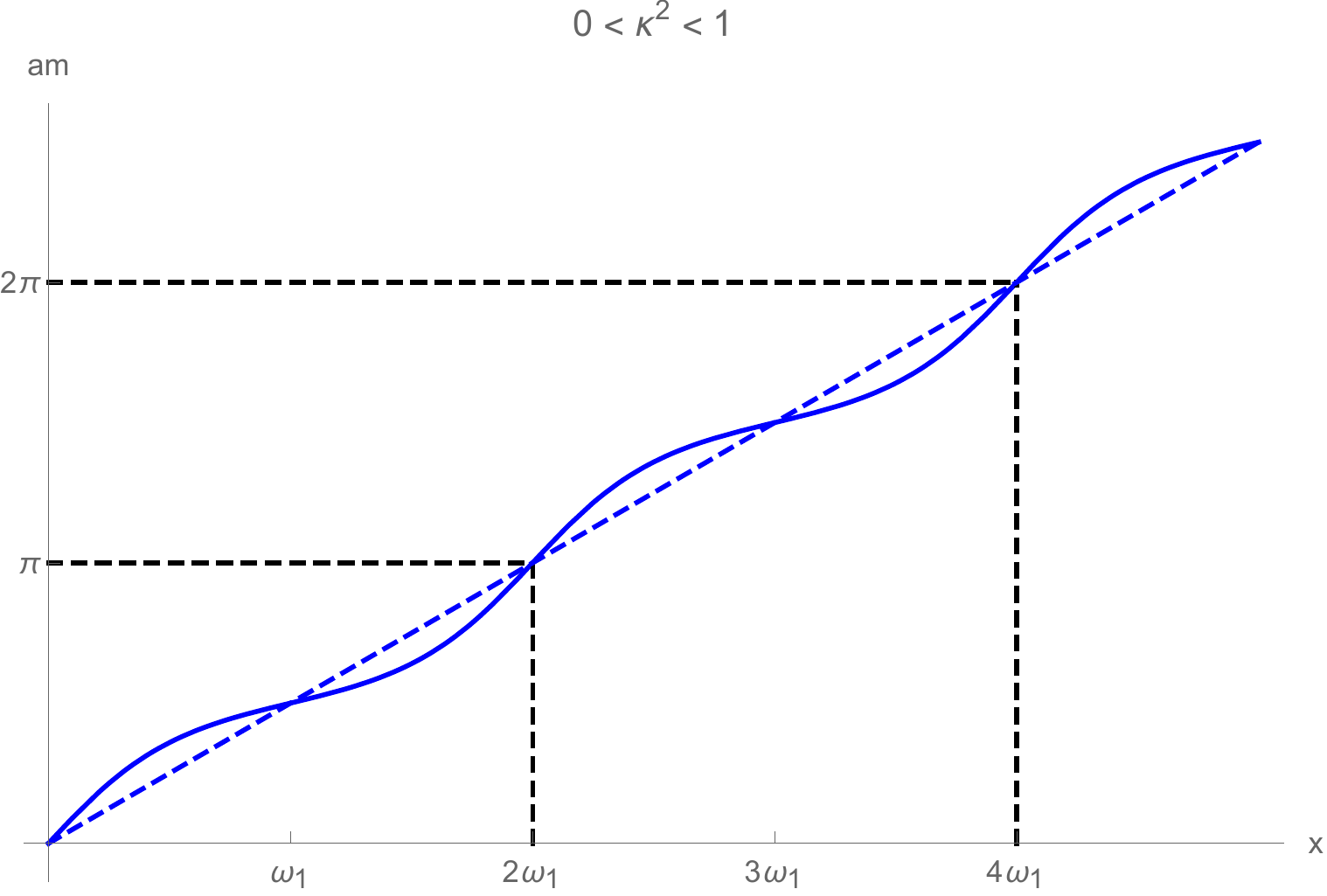}
\caption{The Jacobi amplitude for $\kappa^2=1.25$ (left panel), $\kappa^2=1$ (middle panel) and  $\kappa^2=0.95$ (right panel). The dashed horizontal lines on the left panel mark the extremal values of the function. These values are obtained periodically. On the contrary, on the middle panel, the extremal values are obtained asymptotically. Finally, the dashed blue line of the right panel is the average value of the function over a large number of periods. The plots on the left and right panel clearly show the periodic and quasi-periodic behaviour of the function, respectively.}
\label{fig:Jacobi_Amplitude}
\end{figure}

Elliptic functions are doubly periodic on the complex plane. The elliptic functions appearing in the solutions derived in this work have one real and one imaginary period. These periods form a lattice on the complex plane. Denoting the real half-period as $\omega_1$, it turns out that
\begin{equation}\label{eq:w1_definition}
\omega_1=\begin{cases}
K(\kappa^2),\qquad 0\leq \kappa^2\leq 1\\
\frac{K(\kappa^{-2})}{\kappa},\qquad 1\leq \kappa^2
\end{cases},
\end{equation}
where $K(m)$ is the complete elliptic integral of the first kind, defined as
\begin{equation}
K(m)=\int_0^{\pi/2}\frac{d\phi}{\sqrt{1-m\sin^2\phi}}.
\end{equation}
If $\kappa^2=1$ the real period diverges. The first case of \eqref{eq:w1_definition} is the usual definition of the real period, whereas the second one follows from the properties of the complete elliptic integral and corresponds to a modular transformation on the complex plane. Using the above definition, the quasi-periodicity and periodicity of the Jacobi amplitude reads:
\begin{align}
\mathrm{am}\left(x+2\omega_1\vert \kappa^2\right)&=\mathrm{am}\left(x\vert \kappa^2\right)+\pi,\qquad 0\leq \kappa^2\leq 1,\\
\mathrm{am}\left(x+4\omega_1\vert \kappa^2\right)&=\mathrm{am}\left(x\vert \kappa^2\right),\phantom{+\pi,}\qquad 1\leq \kappa^2.
\end{align}
Figure \ref{fig:Jacobi_Amplitude} shows indicative examples of the different kinds of behaviour of the Jacobi amplitude.

\end{document}